\documentclass[a4paper,11pt]{article}
\usepackage{amsmath,amsfonts,amsthm}
\usepackage{euscript}
\usepackage{mathbbol}
\usepackage{natbib}
\usepackage[thinlines,thiklines]{easybmat}
\begin{document}
\title{On Asymptotic Quantum Statistical Inference}
\author{Richard D. Gill\thanks{URL: {\tt www.math.leidenuniv.nl/$\sim$gill}. Mathematical Institute, Leiden University, The Netherlands} and
   M\u{a}d\u{a}lin I. Gu\c{t}\u{a}\thanks{URL: {\tt www.maths.nottingham.ac.uk/personal/pmzmig/}. School of Mathematical Sciences, University of Nottingham, United Kingdom}}
\date{15 January, 2011}

\newtheorem{theorem}{Theorem}
\newtheorem{question}{Question}
\newtheorem{fact}{Fact to remember}
\newtheorem{definition}{Definition}
\newtheorem{lemma}{Lemma}
\maketitle

\begin{abstract}
We study asymptotically optimal statistical inference 
concerning the unknown state of 
$N$ identical quantum systems, 
using two complementary approaches: a ``poor man's approach''
based on the van Trees inequality, and a rather more sophisticated approach using
the recently developed quantum form of LeCam's theory of Local Asymptotic Normality.
\end{abstract}

\section{Introduction}

The aim of this paper is to show the rich possibilities for 
asymptotically optimal statistical inference for
``quantum i.i.d.\ models''. Despite the possibly exotic context, mathematical statistics
has much to offer, and much that we have leant -- in particular through Jon Wellner's work 
in semiparametric models and nonparametric maximum likelihood estimation --
can be put to extremely good use. Exotic? In today's quantum information engineering,
measurement and estimation schemes are put to work to recover
the state of a small number of quantum states, engineered by the physicist in
his or her laboratory. New technologies are winking at us on the horizon. 
So far, the physicists are largely re-inventing statistical wheels themselves. 
We think it is a pity statisticians are not more involved. 
If Jon is looking for some new challenges... ?

In this paper we do theory. We suppose that one has $N$ 
copies of a quantum system each in the same state depending on
an unknown vector of parameters $\theta$, and one wishes to estimate $\theta$,
or more generally a vector function of the parameters $\psi(\theta)$,
by making some measurement on the $N$ systems together. 
This yields data whose distribution depends on $\theta$ and on the choice of 
the measurement. Given the measurement, we therefore have a 
classical parametric statistical model, though not necessarily an i.i.d.\
model, since we are allowed to bring the $N$ systems together
before measuring the resulting joint system as one quantum object.
In that case the resulting data need not consist of (a function of)
$N$ i.i.d.\ observations, and a key quantum feature is that we
can generally extract more information about $\theta$ using
such ``collective'' or ``joint'' measurements than when we measure
the systems separately.
What is the best we can do as $N\to\infty$, when we are allowed to
optimize both over the measurement and over the ensuing
data-processing?

A statistically motivated, approach to deriving methods
with good properties for large $N$ is to choose the measurement to 
optimize the Fisher information in the data, leaving it to the statistician to
process the data efficiently, using for instance maximum
likelihood or related methods, including Bayesian. This heuristic 
principle has already been shown to work in a number of special 
cases in quantum statistics. Since the measurement
maximizing the Fisher information typically
depends on the unknown parameter value this often has to
be implemented in a two-step approach, first using a small 
fraction of the $N$ systems to get a first approximation to
the true parameter, and then optimizing on the remaining
systems using this rough guess.

The approach favoured by many physicists, on the other hand, is 
to choose a prior distribution and loss function on grounds 
of symmetry and physical interpretation, and then to \emph{exactly}
optimize the Bayes risk over all measurements and estimators, 
for any given $N$.
This approach succeeds in producing attractive methods
on those rare occasions 
when a felicitous combination of all the mathematical ingredients
leads to an analytically tractable solution.

Now it has been observed in a number of problems that
the two approaches result in asymptotically equivalent
estimators, though the measurement schemes can be
strikingly different. Heuristically, this can be understood to follow
from the fact that, in the physicists' approach, for large $N$ 
the prior distribution should become increasingly irrelevant
and the Bayes optimal estimator close to the maximum
likelihood estimator.  Moreover, we expect those estimators
to be asymptotically normal with variances corresponding to
inverse Fisher information. 

Here we link the two approaches by
deriving an asymptotic lower bound on the Bayes risk
of the physicists' approach, in terms of the optimal
Fisher information of the statisticians' approach.
Sometimes one can find in this way  asymptotically
optimal solutions which are much easier to implement
than the exactly optimal solution of the physicists' approach.
On the other hand, it also suggests 
that the physicists' approach, when successful, leads to procedures
which are \emph{asymptotically} optimal for other prior distributions,
and other loss functions, than those used in the computation.
It also suggests that these solutions are 
asymptotically optimal in a pointwise rather than
a Bayesian sense.

In the first part of our paper, we derive our new bound
by combining an existing quantum Cram\'er-Rao bound \citep{holevo82}
with the van Trees inequality, a Bayesian Cram\'er-Rao
bound from classical statistics (\citealp{vantrees68,
gilllevit95}). The former can be interpreted
as a bound on the Fisher information in an arbitrary
measurement on a quantum system, the latter is a bound
on the Bayes risk (for a quadratic loss function) in terms
of the Fisher information in the data. This part of the paper 
can be understood without any familiarity with quantum statistics. 
Applications are given in an appendix to an eprint version of the
paper at arXiv.org.

The paper contains only a brief summary of ``what is a quantum
statistical model''; for more information the reader is referred to the papers
of \citet{barndorffnielsengilljupp03}, and \citet{gill01}.
For an overview of the ``state of the art'' in quantum asymptotic
statistics see \citet{hayashi05} which reprints papers of many authors 
together with introductions by the editor.

After this ``simplistic'' part of the paper we present some of
the recently developed theory of quantum Local Asymptotic Normality 
(also mentioning a number of open problems).
This provides an alternative but more sophisticated route to getting 
asymptotic optimality results, but at the end of the day it also 
explains ``why'' our simplistic approach does indeed work.
In classical statistics, we have learnt to understand asymptotic 
optimality of maximum likelihood estimation through the
idea that an i.i.d.\ parametric model can be closely approximated, locally, by a 
Gaussian shift model with the same information matrix. To say the same thing
in a deeper way, the two models have the same geometric structure of the score functions of
one-dimensional sub-models; and in the i.i.d.\ case, after local rescaling, those 
score functions are asymptotically Gaussian. 

Let us first develop enough notation to
state the main result of the paper and compare it with the
comparable result from classical statistics. Starting on
familiar ground with the latter, suppose we
want to estimate a function $\psi(\theta)$ of a
parameter $\theta$, both represented by real column vectors
of possibly different dimension, based on $N$ i.i.d.\ observations from
a distribution with Fisher information matrix $I(\theta)$.
Let $\pi$ be a prior density on the parameter space and
let $\widetilde G(\theta)$ be a symmetric positive-definite matrix
defining a quadratic loss function 
$l(\widehat\psi^{(N)},\theta)= (\widehat\psi^{(N)}-\psi(\theta))^\top \widetilde G(\theta) (\widehat\psi^{(N)}-\psi(\theta))$. 
(Later we will use $G(\theta)$, without the tilde, in the special case when $\psi$ is $\theta$ itself).
Define the mean square error matrix 
$V^{(N)}(\theta)=\mathbb E_\theta (\widehat\psi^{(N)}-\psi(\theta)) (\widehat\psi^{(N)}-\psi(\theta))^\top$
so that the risk can be written $R^{(N)}(\theta)=\mathrm{trace}\,\widetilde G(\theta)V^{(N)}(\theta)$.
The Bayes risk is $R^{(N)}(\pi)=\mathbb E_\pi \mathrm{trace}\,\widetilde GV^{(N)}$.
Here, $\mathbb E_\theta$ denotes expectation over the data for given $\theta$,
$\mathbb E_\pi$ denotes averaging over $\theta$ with respect to the prior $\pi$.
The estimator $\widehat\psi^{(N)}$ is completely arbitrary.
We assume the prior density to be smooth, compactly supported 
and zero on the smooth boundary of its support. Furthermore a certain
quantity roughly interpreted as ``information in the prior'' 
must be finite.
Then it is very easy to show \citep{gilllevit95}, using the van Trees inequality,
 that under
minimal smoothness conditions on the statistical model,
\begin{equation}\label{mainresult1}
\mathop{\lim\inf}\limits_{N\to\infty}N R^{(N)}(\pi) ~\ge ~\mathbb E_\pi \mathrm{trace} \, G I^{-1}
\end{equation}
where $G=\psi' \widetilde G\psi'^\top$ and $\psi'$ is the matrix of
partial derivatives of elements of $\psi$ with respect to those of $\theta$.

Now in quantum statistics the data depends on the choice of measurement 
and the measurement should be tuned to the loss function.
Given a measurement $M^{(N)}$ on $N$ copies of
the quantum system, denote by $\overline I_M^{(N)}$ the average
Fisher information (i.e., Fisher information divided by $N$) in the
data. The \citet{holevo82} quantum Cram\'er-Rao bound, as 
extended by \citet{hayashimatsumoto04} to the quantum i.i.d.\ model, 
can be expressed as saying that, for all $\theta$, $G$, $N$ and  $M^{(N)}$,
\begin{equation}\label{holevo}
\mathrm{trace} \, G(\theta) (\overline I_M^{(N)}(\theta))^{-1}~\ge~\EuScript C_G(\theta)
\end{equation}
for a certain quantity $\EuScript C_G(\theta)$,
which depends on the specification of the quantum statistical model
(state of one copy,  derivatives of the state with respect to parameters, 
and loss function $G$)  \emph{at the point $\theta$} only, 
i.e., on local or pointwise model features (see (\ref{defCG1}) below).

We aim to prove that under minimal
smoothness conditions on the quantum statistical model, and conditions
on the prior similar to those needed in the classical case, 
but under essentially no conditions on the estimator-and-measurement sequence,
\begin{equation}\label{mainresult}
\mathop{\lim\inf}\limits_{N\to\infty}N R^{(N)}(\pi) ~\ge ~\mathbb E_\pi \EuScript C_G
\end{equation}
where, as before, $G=\psi' \widetilde G\psi'^\top$.
The main result (\ref{mainresult}) is exactly the bound one would hope for, 
from heuristic statistical principles.
In specific models of interest, the right hand side is often easy
to calculate. Various specific measurement-and-estimator sequences,
motivated by a variety of approaches, can also be shown in
interesting examples to achieve the bound, see the appendix to the eprint
version of this paper.

It was also shown in \citet{gilllevit95}, how---in the classical statistical context---one 
can replace a fixed prior $\pi$
by a sequence of priors indexed by $N$, concentrating more and more
on a fixed parameter value $\theta_0$, at rate $1/\sqrt N$. Following their
approach would, in the quantum context, lead to the pointwise asymptotic lower bounds 
\begin{equation}\label{regular}
\mathop{\lim\inf}\limits_{N\to\infty}N R^{(N)}(\theta) ~\ge ~\EuScript C_G(\theta)
\end{equation}
for each $\theta$, for \emph{regular}
estimators, and to local asymptotic minimax bounds 
\begin{equation}\label{minimax}
\lim_{M\to\infty}\mathop{\lim\inf}\limits_{N\to\infty}\sup_{\|\theta-\theta_0\|\le N^{-1/2}M}
N R^{(N)}(\theta) ~\ge ~\EuScript C_G(\theta_0)
\end{equation}
for \emph{all} estimators, but we do not further develop that
theory here.
In classical statistics the theory of Local Asymptotic Normality is the way
to unify, generalise, and understand this kind of result. In the last section 
of this paper we introduce the now emerging quantum generalization of this theory.

The basic tools used in the first part of this paper have now all been
mentioned, but as we shall see, the proof is not a routine
application of the van Trees inequality.
The missing ingredient will be provided by  the following new \emph{dual} 
bound to
(\ref{holevo}):
for all $\theta$, $K$, $N$ and  $M^{(N)}$,
\begin{equation}\label{dualholevo}
\mathrm{trace}\, K(\theta) \overline I_M^{(N)}(\theta)~\le~\EuScript C^K(\theta)
\end{equation}
where $\EuScript C^K(\theta)$ actually equals $\EuScript C_G(\theta)$ 
for a certain $G$ defined in terms of $K$ (as explained in
Theorem \ref{dualholevothm} below).
This is an \emph{upper} bound on Fisher information, in contrast to
(\ref{holevo}) which is a lower bound on inverse Fisher information.
The new inequality (\ref{dualholevo})
follows from the convexity of the sets of information matrices 
and of inverse information matrices  for arbitrary measurements 
on a quantum system, and these convexity
properties have a simple statistical explanation.
Such dual bounds have cropped
up incidentally in quantum statistics, for instance in \citet{gillmassar00},
but this is the first time a connection is established.

The argument for (\ref{dualholevo}), and given that, for (\ref{mainresult}), 
is based on some general structural features of
quantum statistics, and hence it is not necessary to be familiar with
the technical details of the set-up. 

In the next section we will summarize the
i.i.d.\ model in quantum statistics, focussing  on the
key facts which will be used in the proof of the dual Holevo bound 
(\ref{dualholevo}) and of our main result, the
asymptotic lower bound (\ref{mainresult}).

These proofs are given in a subsequent section, where
no further ``quantum''  arguments will be used. 

In the final section we will show how the bounds correspond
to recent results in the theory of Q-LAN, according to which the i.i.d.\ 
model converges to a quantum Gaussian shift experiment, with the same
Holevo bounds, which are actually attainable in the Gaussian case.
An eprint version of this paper, Gill and Gu\c{t}\u{a} (2012) includes an appendix with some worked
examples.

\section{Quantum statistics: the i.i.d.\ parametric case.}

The basic objects in quantum statistics are \emph{states} and
\emph{measurements}, defined in terms of certain 
operators on a complex Hilbert space. 
To avoid technical complications we restrict attention to the 
finite-dimensional case, already rich in structure and
applications, when operators are represented by
ordinary (complex) matrices.

\paragraph{States and measurement}

The state of a $d$-dimensional system is represented by a $d\times d$
matrix $\rho$, called the \emph{density matrix} of the state, 
having the following properties: $\rho^*=\rho$ (self-adjoint
or Hermitian), $\rho\ge \mathbf 0$ (non-negative), $\mathrm{trace}(\rho)=1$
(normalized).  ``Non-negative'' actually implies ``self-adjoint'' but it does no harm to 
 emphasize both properties.  $\mathbf 0$ denotes the zero matrix; $\mathbf 1$ will denote 
 the identity matrix. 
 \medskip
 \\
 \noindent \emph{Example}: when $d=2$, every density matrix can be written
in the form $\rho=\frac12(\mathbf 1 + \theta_1 \sigma_1+\theta_2\sigma_2
+\theta_3 \sigma_3)$ where 
\begin{equation*}
\sigma_1=\left(\begin{matrix} 0& 1\\ 1& 0
\end{matrix}\right),\quad\sigma_2=\left(\begin{matrix} 0 &-i\\ i& 0
\end{matrix}\right),\quad\sigma_3=\left(\begin{matrix} 1 &0\\ 0 &-1
\end{matrix}\right)
\end{equation*}
 are the three Pauli matrices and where $\theta_1^2+\theta_2^2+\theta_3^2\le 1$. \hfill$\qed$
  \medskip
\\
  ``Quantum statistics'' concerns the situation when 
 the state of the system  $\rho(\theta)$ depends on a (column) vector $\theta$
 of $p$ unknown 
 (real) parameters.
  \medskip
 \\
 \emph{Example}: a completely unknown two-dimensional
 quantum state depends on a vector of three real parameters, 
 $\theta=(\theta_1,\theta_2,\theta_3)^\top$, known to lie in the unit ball.
 Various interesting submodels can be described geometrically: e.g.,
 the equatorial plane; the surface of the ball; a straight line through the origin.
 More generally, a completely unknown $d$-dimensional state depends on
 $p=d^2-1$ real parameters.\hfill $\qed$
  \medskip
 \\
\emph{Example}: in the previous example the two-parameter case 
obtained by demanding that  $\theta_1^2+\theta_2^2+\theta_3^2= 1$
is called the case of a two-dimensional pure state. In general,
a state is called pure if $\rho^2=\rho$ or equivalently $\rho$ has
rank one. A completely unknown pure $d$-dimensional state depends
on $p=2(d-1)$ real parameters.\hfill $\qed$
  \medskip
 
A measurement on a quantum system is characterized by the outcome space,
which is just a measurable space $(\EuScript X,\EuScript B)$, and a \emph{positive operator
valued measure} (POVM) $M$ on this space. This means that for each $B\in\EuScript B$  there corresponds a $d\times d$ non-negative self-adjoint matrix $M(B)$, 
together having the usual properties
 of an ordinary  (real) measure (sigma-additive), with moreover $M(\EuScript X)=\mathbf 1$.
 The probability distribution of the outcome of doing measurement $M$ on state $\rho(\theta)$
 is given by the Born law, or trace rule: $\Pr(\textrm{outcome}\in B)=\mathrm{trace}(\rho(\theta)M(B))$.
 It can be seen that this is indeed a bona-fide probability distribution on the sample
 space $(\EuScript X,\EuScript B)$. Moreover it has a density with respect to
 the finite real measure $\mathrm{trace}(M(B))$. 
 \medskip
 \\
\emph{Example}: the most simple measurement is defined by choosing an orthonormal
basis of $\mathbb C^d$, say $\psi_1$,\dots,$\psi_d$, taking the outcome space
to be the discrete space $\EuScript X=\{1,\dots,d\}$, and defining $M(\{x\})=\psi_x\psi_x^*$
for $x\in\EuScript X$;
or in physicists' notation, $M(\{x\})=|\psi_x\rangle\langle\psi_x|$.
One computes that $\Pr(\textrm{outcome}=x)=\psi_x^* \rho(\theta) \psi_x =
\langle \psi_x|\rho|\psi_x\rangle$. If the state is pure then $\rho=\phi\phi^*=
|\phi\rangle\langle\phi|$ for some $\phi=\phi(\theta)\in\mathbb C^d$ of length $1$ and depending
on the parameter $\theta$. One finds that  $\Pr(\textrm{outcome}=x)=|\psi_x^*\phi|^2
=|\langle\psi_x|\phi\rangle|^2$.   \hfill $\qed$
  \medskip

So far we have discussed state and measurement for a single quantum system.
This encompasses also the case of $N$ copies of the system,
via a tensor product construction, which we will now summarize.
The joint state of $N$ identical copies of a single system having state $\rho(\theta)$ is
$\rho(\theta)^{\otimes N}$, a density matrix on a space of dimension $d^N$. 
A joint or collective measurement on these systems is specified by a POVM on this
large  tensor product Hilbert space. An important point
is that joint measurements give many more possibilities than measuring the separate
systems independently, or even measuring the separate systems adaptively.

 \begin{fact} State plus measurement determines probability distribution of
 data.
 \end{fact}

\paragraph{Quantum Cram\'er-Rao bound.}

Our main input is going to be the \citet{holevo82} quantum Cram\'er-Rao bound, 
with its extension to the i.i.d.\ case due to \citet{hayashimatsumoto04}. 

Precisely because of quantum phenomena, different measurements, incompatible
with one another, are appropriate when we are interested in different components
of our parameter, or more generally, in different loss functions. The bound
concerns estimation of  $\theta$ itself rather than a function thereof, and depends on
a quadratic loss function defined by a symmetric real non-negative matrix $G(\theta)$ which 
may depend on the actual parameter value $\theta$. 
For a given estimator $\widehat\theta^{(N)}$ computed from the outcome
of some measurement $M^{(N)}$ on $N$ copies of our system, define its 
mean square error matrix $V^{(N)}(\theta)=\mathbb E_\theta (\widehat\theta^{(N)}-\theta)
(\widehat\theta^{(N)}-\theta)^\top$. The risk function when
using the quadratic loss determined by $G$ is 
$R^{(N)}(\theta)=\mathbb E_\theta (\widehat\theta^{(N)}-\theta)^\top G(\theta)  (\widehat\theta^{(N)}-\theta)=\mathrm{trace}(G(\theta)V^{(N)}(\theta))$. 

One may expect the risk of good measurements-and-estimators to decrease 
like $N^{-1}$ as $N\to \infty$.
The quantum Cram\'er-Rao bound confirms that this is the best
rate to hope for: it states that for unbiased estimators 
of a $p$-dimensional parameter $\theta$, based on
arbitrary joint measurements on $N$ copies, 
\begin{equation}\label{defCG1}
N R^{(N)}(\theta) ~\ge~  \EuScript C_G(\theta) ~=
~\inf_{\vec X,V:V \ge Z(\vec X)} \mathrm{trace}(G(\theta) V)
\end{equation}
where $\vec X = (X_1,\dots,X_p)$, the $X_i$ are
$d\times d$ self-adjoint matrices satisfying  
\begin{equation}\label{eq.unbiasedness}
\partial/\partial\theta_i\, \mathrm{trace}(\rho(\theta) X_j)=\delta_{ij},
\end{equation}
$Z$ is the $p\times p$ self-adjoint matrix with elements $\mathrm{trace}(\rho(\theta)X_i X_j)$, 
and $V$ is a real symmetric matrix.
It is possible to solve the optimization over $V$ for given $\vec X$ leading
to the formula
\begin{equation}\label{defCG2}
\EuScript C_G(\theta) ~=
~\inf_{\vec X} \mathrm{trace}\bigl( \Re (G^{1/2}Z(\vec X)G^{1/2} ) 
               + \mathrm{abs}\Im (G^{1/2}Z(\vec X )G^{1/2})\bigr)
 \end{equation}
where $G=G(\theta)$.
The absolute value of a matrix is found by diagonalising it and taking absolute values of
the eigenvalues. 
We'll assume that the bound is finite, i.e., there exists
$\vec X$ satisfying the constraints. A sufficient condition for this is that the
Helstrom quantum information matrix $H$ introduced in (\ref{helstrom})
below is nonsingular.

For specific interesting models, it often turns out not difficult to compute the bound
$\EuScript C_G(\theta)$. Note, it is a bound which depends only on the  
density matrix of one system ($N=1)$ and its derivative
with the respect to the parameter, and on the loss function, both at the given point $\theta$.
It can be found by solving a finite-dimensional optimization problem.

We will not be concerned with the specific form of the bound. What we are going to
need, are just two key properties. 

Firstly:  the bound is local, and applies to the larger
class of \emph{locally unbiased estimators}. This means to say that 
\emph{at the given point $\theta$}, $\mathbb E_\theta \widehat \theta^{(N)}=\theta$,
and at this point also $\partial/\partial\theta_i\, \mathbb E_\theta \widehat \theta_j^{(N)}=\delta_{ij}$.
Now, it is well known that the ``estimator'' $\theta_0+I(\theta_0)^{-1}S(\theta_0)$,
where $I(\theta)$ is Fisher information and $S(\theta)$ is score function, is locally
unbiased at $\theta=\theta_0$ \emph{and achieves the Cram\'er-Rao bound there}.
Thus the Cram\'er-Rao bound for \emph{locally} unbiased estimators is sharp.
Consequently, we can rewrite the  bound (\ref{defCG1}) in the form (\ref{holevo})
announced above,
where $\overline I_M^{(N)}(\theta)$ is the \emph{average} (divided by $N$) Fisher
information in the outcome of an arbitrary measurement $M=M^{(N)}$ on $N$ copies
and the right hand side is defined in (\ref{defCG1}) or (\ref{defCG2}).

\begin{fact} We have a family of computable lower bounds  
on the inverse average Fisher information matrix for an 
arbitrary measurement on $N$ copies, given by (\ref{holevo}) and (\ref{defCG1})
or (\ref{defCG2}),
\end{fact}

Secondly, for given $\theta$, define the following two sets of
positive-definite symmetric real matrices, in one-to-one correspondence
with one another through the mapping ``matrix inverse''. 
The matrices $G$ occurring in the definition are also 
taken to be positive-definite symmetric real.
\begin{equation}\label{defV}
\EuScript V = \{V:\mathrm{trace}(G V) \ge \EuScript  C_G~ \forall ~ G\},
\end{equation}
\begin{equation}\label{defI}
\EuScript I = \{I:\mathrm{trace}(G I^{-1}) \ge  \EuScript  C_G~ \forall ~ G\}.
\end{equation}

Elsewhere (Gill, 2005) we have given a proof by matrix algebra
that that the set $\EuScript I$ is convex (for $\EuScript V$, convexity is obvious),
and that the inequalities defining $\EuScript V$ define supporting hyperplanes 
to that convex set, i.e., all the inequalities are achievable in $\EuScript V$, 
or equivalently $\EuScript  C_G=\inf_{V\in\EuScript V} \mathrm{trace}(G V)$. But now, with the
tools of Q-LAN behind us (well -- ahead of us -- see the last section of this paper),
we can give a short, statistical, explanation which is simultaneously a short, complete, proof.

The quantum statistical problem of collective measurements
on $N$ identical quantum systems, when rescaled at the proper $\sqrt{\vphantom{|}}N$-rate, 
approaches a quantum Gaussian
problem as $N\to\infty$, as we will see the last section of this paper.
In this problem, $\EuScript V$ consists precisely of all the covariance 
matrices of locally unbiased estimators achievable (by suitable choice of measurement)
in the limiting  $p$-parameter quantum Gaussian statistical model.
The inequalities defining $\EuScript V$ are exactly the Holevo bounds
for that model, and each of those bounds, as we show in Section 4, 
is attainable. Thus, for each $G$, 
there exists a $V\in\EuScript V$ achieving equality in 
$\mathrm{trace}(G V) \ge \EuScript  C_G$. 
It follows from this that $\EuScript I$ consists of all non-singular
information matrices
(augmented with all non-singular matrices smaller than an information matrix) 
achievable by choice of measurement on the same quantum Gaussian model. 
Consider the set of information 
matrices attainable by some measurement, together
with all smaller matrices; and consider the set of variance matrices of locally
unbiased estimators based on arbitrary measurements, together with all larger matrices. 
Adding zero
mean noise to a locally unbiased estimator preserves its local unbiasedness,
so adding larger matrices to the latter set does not change it, by the mathematical definition
of measurement, which includes addition of outcomes of arbitrary auxiliary randomization.
The set of information matrices is convex: choosing measurement $1$
with probability $p$ and measurement $2$ with probability $q$ while
remembering your choice, gives a measurement whose Fisher information
is the convex combination of the informations of measurements $1$ and $2$.
Augmenting the set with all matrices smaller than something in the set, preserves
convexity. The set of variances of locally unbiased estimators is convex, 
by a similar randomization argument. Putting this together, we obtain

\begin{fact}
For given $\theta$, 
both $\EuScript V$
and $\EuScript I$ defined in (\ref{defV}) and (\ref{defI}) are convex,
and all the inequalities defining these sets are achieved by points
in the sets.
\end{fact}

\section{An asymptotic Bayesian information bound}

We will now introduce the van Trees inequality, a Bayesian Cram\'er-Rao
bound, and combine it with the Holevo bound (\ref{holevo}) via
derivation of a dual bound following from the convexity of the sets (\ref{defCG1}) 
and (\ref{defCG2}). We return to the problem of estimating the (real, column)  vector function 
$\psi(\theta)$ of the (real, column) vector parameter $\theta$
of a state $\rho(\theta)$ based on collective measurements of $N$ identical copies.
The dimensions of $\psi$ and of $\theta$ need not be the same. The sample
size $N$ is largely suppressed from the notation.
Let $V$ be the mean square error matrix of an arbitrary estimator $\widehat\psi$,
thus $V(\theta) = \mathbb E_\theta (\widehat\psi-\psi(\theta))(\widehat\psi-\psi(\theta))^\top$.
Often, but not necessarily, we'll have $\widehat\psi=\psi(\widehat\theta)$ for some
estimator of $\theta$.
Suppose we have a quadratic loss function 
$(\widehat\psi-\psi(\theta))^\top \widetilde G(\theta) (\widehat\psi-\psi(\theta))$
where $\widetilde G$ is a positive-definite matrix function of $\theta$, then
the Bayes risk with respect to a given prior $\pi$
can be written  $R(\pi)=\mathbb E_\pi \mathrm{trace}\, \widetilde G V$.
We are going to prove the following theorem:

\begin{theorem}
Suppose $\rho(\theta):\theta\in\Theta\subseteq \mathbb R^p$
is a smooth quantum statistical model and suppose
$\pi$ is a smooth prior density on a compact subset $\Theta_0\subseteq \Theta$, 
such that $\Theta_0$ has a piecewise smooth boundary, on which $\pi$ is zero.  
Suppose moreover the quantity $\EuScript J(\pi)$ defined in (\ref{defJpi}) below, is finite. Then
\begin{equation}\label{mainresult2}
\mathop{\lim\inf}\limits_{N\to\infty}N R^{(N)}(\pi) ~\ge ~\mathbb E_\pi \EuScript C_{G_0}
\end{equation}
where $G_0=\psi' \widetilde G\psi'^\top$ (and assumed to be positive-definite), 
$\psi'$ is the matrix of
partial derivatives of elements of $\psi$ with respect to those of $\theta$,
and $\EuScript C_{G_0}$  is defined by (\ref{defCG1})  or  (\ref{defCG2}).
\end{theorem}
``Once continuously differentiable'' is enough smoothness. Smoothness of
the quantum statistical model implies smoothness of the classical statistical
model following from applying an arbitrary measurement to $N$ copies of
the quantum state. Slightly weaker but more elaborate smoothness 
conditions on the statistical model and prior are spelled out
in \citet{gilllevit95}. The restriction that $G_0$ be non-singular
can probably be avoided by a more detailed analysis.

Let $\overline I_M$ denote the average Fisher information matrix
for $\theta$ based on a given collective measurement on the $N$ copies. Then the
 van Trees inequality states that for all matrix functions $C$ of $\theta$, of size
 $\mathrm {dim}(\psi)\times\mathrm{dim}(\theta)$, 
  \begin{equation}\label{e:vt}
 N\mathbb E_\pi  \mathrm{trace}\, \widetilde G V ~\ge~
 \frac{(\mathbb E_\pi\mathrm{trace}\, C\psi'^\top)^2}
 {\mathbb E_\pi \mathrm{trace}\, \widetilde G^{-1} C \overline I_M C^\top +\frac 1 N  
 \mathbb E_\pi \frac {(C\pi)'^\top \widetilde G^{-1} (C\pi)'} {\pi^2} }
 \end{equation}
 where the primes in $\psi'$ and in $(C\pi)'$ both denote differentiation,
 but in the first case converting the vector $\psi$ into the matrix of
 partial derivatives of elements of $\psi$ with respect to elements of
 $\theta$, of size $\mathrm {dim}(\psi)\times\mathrm{dim}(\theta)$,
 in the second case converting the matrix $C\pi$ into the
 column vector, of the same length as $\psi$,
 with row elements $\sum_j (\partial/\partial\theta_j)(C\pi)_{ij}$.
To get an optimal bound we need to choose $C(\theta)$ cleverly.
 
 First though,  note that the Fisher information appears in the denominator
 of the van Trees bound. This is a nuisance since we have a 
 Holevo's  \emph{lower}
 bound (\ref{holevo}) to the \emph{inverse} Fisher information.
 We would like to have an \emph{upper} bound on the
 information itself, say of the form (\ref{dualholevo}),
 together with a recipe for computing $\EuScript C^K$.
 
 All this can be obtained from the convexity of the sets
 $\EuScript I$ and $\EuScript V$ defined in (\ref{defI})
 and (\ref{defV}) and the non-redundancy of the
inequalities appearing in their definitions.
 Suppose $V_0$ is a boundary point of
$\EuScript V$. Define $I_0=V_0^{-1}$.  
Thus $I_0$ (though not necessarily an attainable average
information matrix $\overline I_M^{(N)}$) satisfies the Holevo
bound for each positive-definite $G$, and attains equality in one 
of them, say with $G=G_0$.  In the language of convex sets,
and ``in the $V$-picture'',
$\mathrm{trace}\, G_0 V= \EuScript C_{G_0}$ is a 
supporting hyperplane to $\EuScript V$ at $V=V_0$.

Under the mapping ``matrix-inverse" the hyperplane
$\mathrm{trace}\, G_0 V= \EuScript C_{G_0}$ in the $V$-picture 
maps to the smooth surface 
$\mathrm{trace}\, G_0 I^{-1}= \EuScript C_{G_0}$
touching the set $\EuScript I$ at $I_0$ in the $I$-picture.
Since $\EuScript I$ is convex, the tangent plane
to the smooth surface at $I=I_0$ must be 
a supporting hyperplane to $\EuScript I$
at this point.
The matrix derivative of the operation of matrix inversion
 can be written $\mathrm d A^{-1}/\mathrm d x= - A^{-1} (\mathrm d A / \mathrm d x) A^{-1}$.
This tells us that the equation of the tangent plane is
$\mathrm{trace}\, G_0 I_0 ^{-1} I I_0^{-1} = \mathrm{trace} \, G_0 I_0^{-1} =\EuScript C_{G_0}$.
Since this is simultaneously a supporting hyperplane to $\EuScript I$ we
deduce that for all $I\in \EuScript I$, 
$\mathrm{trace}\, G_0 I_0 ^{-1} I I_0^{-1} \le \EuScript C_{G_0}$. Defining
$K_0=I_0^{-1}G_0 I_0 ^{-1}$ and $\EuScript C^{K_0}=\EuScript C_{G_0}$ 
we rewrite this inequality as $\mathrm{trace}\, K_0 I \le \EuScript C^{K_0}$.

A similar story can be told when we start in the $I$-picture with a supporting
hyperplane (at $I=I_0$) to $\EuScript I$ of the form $\mathrm{trace}\, K_0 I = \EuScript C^{K_0}$
for some symmetric positive-definite $K_0$. It maps to the smooth surface
$\mathrm{trace}\, K_0 V^{-1} = \EuScript C^{K_0}$, with tangent plane
$\mathrm{trace}\, K_0 V_0^{-1} I V_0^{-1} = \EuScript C^{K_0}$ at $V=V_0=I_0^{-1}$.
By strict convexity of the function ``matrix inverse'', the tangent plane touches the
smooth surface only at the point $V_0$. Moreover, the smooth surface lies
above the tangent plane, but below $\EuScript V$. This makes $V_0$ the
unique minimizer of $\mathrm{trace}\, K_0 V_0^{-1} I V_0^{-1}$ in $\EuScript V$.

It would be useful to extend these computations to allow
singular $I$, $G$ and $K$.
Anyway, we summarize what we have so far in a theorem.
 \begin{theorem}\label{dualholevothm}
Dual to the Holevo family of lower bounds on average inverse
 information, $\mathrm{trace}\, G \overline I_M^{-1}\ge\EuScript C_G$ for each 
 positive-definite $G$, 
 we have a family of upper bounds on information, 
 \begin{equation}\label{dualholevo2}
 \mathrm{trace}\, K \overline I_M\le\EuScript C^K~~\text{for each}~~K.
 \end{equation}
 If $I_0\in\EuScript I$ satisfies $\mathrm{trace} \,G_0 I_0^{-1} = \EuScript C_{G_0}$
then with $K_0=I_0^{-1}G_0 I_0 ^{-1}$, $\EuScript C^{K_0}=\EuScript C_{G_0}$.
Conversely if $I_0\in\EuScript I$ satisfies $\mathrm{trace} \,K_0 I_0 = \EuScript C^{K_0}$
then with $G_0=I_0 K_0 I_0$, $\EuScript C_{G_0}=\EuScript C^{K_0}$. Moreover,
none of the bounds is redundant, in the sense that for all positive-definite
$G$ and $K$,
$\EuScript  C_G=\inf_{V\in\EuScript V} \mathrm{trace}(G V)$ and 
$\EuScript  C^K=\sup_{I\in\EuScript I} \mathrm{trace}(K I)$. 
The minimizer in the first equation is unique.
\end{theorem}

 Now we are ready to apply the van Trees inequality. First we make a guess for what the
 left hand side of (\ref{e:vt}) should look like, at its best. Suppose we use an estimator
 $\widehat\psi=\psi(\widehat\theta)$ where $\widehat\theta$ makes optimal
 use of the information in the measurement $M$. Denote
 now by $I_M$ the asymptotic normalized Fisher information of a sequence
 of measurements. Then we expect that the asymptotic normalized
 covariance matrix $V$
 of $\widehat\psi$ is equal to $\psi' I_M^{-1}\psi'^\top$ and therefore
 the asymptotic normalized Bayes risk should be 
 $\mathbb E_\pi\mathrm{trace}\,\widetilde G \psi' I_M^{-1}\psi'^\top
 =\mathbb E_\pi\mathrm{trace}\,\psi'^\top \widetilde G \psi' I_M^{-1}$. This is bounded
 below by the integrated Holevo bound $\mathbb E_\pi \EuScript C_{ G_0}$ with 
 $ G_0 = \psi'^\top \widetilde G \psi'$.
 Let $I_0\in\EuScript I$ satisfy
$\mathrm{trace}\,  G_0 I_0^{-1} =\EuScript C_{ G_0}$;
its existence and uniqueness are given by Theorem \ref{dualholevothm}.
(Heuristically we expect that $I_0$ is asymptotically attainable).
By the same Theorem, with $K_0=I_0^{-1}G_0 I_0^{-1}$,
 $\EuScript C^{K_0}=\EuScript C_{ G_0}=\mathrm{trace}\, G_0 I_0^{-1}
=\mathrm{trace}\,\psi'^\top \widetilde G \psi' I_0^{-1}$.

Though these calculations are informal, they lead us to try  the matrix function 
$C=\widetilde G \psi' I_0^{-1}$.  Define $V_0=I_0^{-1}$.
With this choice, in the numerator of the van Trees inequality,
we find the square of  $\mathrm{trace}\, C\psi'^\top=\mathrm{trace}\,\widetilde G \psi' I_0^{-1}\psi'^\top
=\mathrm{trace}\, G_0 V_0 =\EuScript C_{ G_0}$. In the main term of the denominator,
we find $\mathrm{trace}\,\widetilde G^{-1}\widetilde G \psi' I_0^{-1}\overline I_M  I_0^{-1} \psi'^\top 
\widetilde G=
\mathrm{trace}\,I_0^{-1} G_0 I_0^{-1} \overline I_M=
\mathrm{trace}\,K_0 \overline I_M \le\EuScript C^{K_0}=\EuScript C_{G_0}$
by the dual Holevo bound (\ref{dualholevo2}).
This makes the numerator of the van Trees bound equal to the square of
this part of the denominator, and using the inequality $a^2/(a+b)\ge a-b$ we find
  \begin{equation}\label{e:vt2}
 N\mathbb E_\pi  \mathrm{trace}\, G V ~\ge~
\mathbb E_\pi \EuScript C_{G_0} -
 \frac 1 N  
 \EuScript J(\pi)
  \end{equation}
where
\begin{equation}\label{defJpi}
\EuScript J(\pi) ~=~\mathbb E_\pi \frac {(C\pi)'^\top \widetilde G^{-1} (C\pi)'} {\pi^2}
\end{equation} 
with
$C=\widetilde G \psi' V_0$ and $V_0$ 
uniquely achieving in $\EuScript V$ 
the  bound $\mathrm{trace}\, G_0 V\ge\EuScript C_{G_0}$, where $ G_0 = \psi'^\top \widetilde G \psi'$.
Finally, provided $\EuScript J(\pi)$ is finite (which depends on the prior distribution
 and on properties of the model), we obtain the asymptotic lower bound
 \begin{equation}\label{e:vtasympt}
\mathop{\lim\inf}\limits_{N\to\infty} N\mathbb E_\pi  \mathrm{trace}\, \widetilde G V ~\ge~
\mathbb E_\pi\EuScript C_{G_0}. 
 \end{equation}

\section{Q-LAN for i.i.d.\  models}
In this section we sketch some elements of a theory of comparison and convergence of quantum statistical models, which is currently being developed in analogy to the LeCam theory of classical statistical models. We illustrate the theory with the example of local asymptotic normality for (finite dimesional) i.i.d.\ quantum states, which provides a route to proving that the Holevo bound is asymptotically achievable.  For more details we refer to the papers \cite{GK,GJK,GJ,KG}, for the i.i.d.\ case and to \cite{G} for the case of  mixing quantum Markov chains. 

The Q-LAN theory surveyed here concerns \emph{strong} local asymptotic normality. 
Just as in the classical case, the ``strong'' version of the theory enables us not only to derive asymptotic bounds, but also to actually construct asymptotically optimal statistical procedures, 
by explicitly lifting the optimal solution of
the asymptotic problem back to the finite $N$ situation, where it is approximately optimal. 
It will be useful to build up theory and applications of the corresponding \emph{weak} local asymptotic normality concept. A start has been made by \cite{GJ}. Such a theory would be easier to apply, and would be sufficient to obtain rigorous asymptotic bounds, but would not contain recipes for how to attain them. At present there are some situations (involving degeneracy) where stong local asymptotic normality is conjectured but not yet proven. It would be interesting to study these analytically tricky problems first using the simpler tools of weak Q-LAN.

\subsection{Convergence of classical statistical models}

To facilitate the comparison between classical and quantum, we will start with a brief summary of some basic notions from the classical theory of convergence of statistical models, specialised to the case of dominated models. 

Recall that if $\mathbb{P}_{\theta}$ is a probability distribution on $(\Omega, \Sigma)$ with $\theta\in \Theta$ unknown, then model $\mathcal{P}= \{ \mathbb{P}_{\theta} :\theta \in \Theta\}$ is called dominated if $\mathbb{P}_{\theta}\ll \mathbb{P}$ for some measure $\mathbb{P}$. We will denote by $p_{\theta}$ the probability density of  $\mathbb{P}_{\theta}$ with respect to $\mathbb{P}$. Similarly, let $\mathcal{P}^{\prime}:=\{ \mathbb{P}_{\theta}^{\prime}:\theta\in \Theta\}$ be another model on $(\Omega^{\prime}, \Sigma^{\prime})$ with densities $p_{\theta}^{\prime}= d\mathbb{P}^{\prime}_{\theta}/d\mathbb{P}^{\prime}$. Then we say that $\mathcal{P}$ and $\mathcal{P}^{\prime}$ are statistically equivalent  (denoted $\mathcal{P}\sim \mathcal{P}^{\prime}$) if their distributions can be transformed into each other via randomisations, i.e., if there exists a linear transformation 
$$
R: L^{1}(\Omega, \Sigma, \mathbb{P})\to L^{1}(\Omega^{\prime}, \Sigma^{\prime}, \mathbb{P}^{\prime})
$$ 
mapping probability densities into probability densities,  such that for all $\theta\in \Theta$
$$
R(p_{\theta}) = p_{\theta}^{\prime},  
$$
and similarly in the opposite direction.  In particular,  $S:\Omega\to \Omega^{\prime}$ is a sufficient statistic for 
$\mathcal{P}$ if and only if $\mathcal{P}\sim \mathcal{P}^{\prime}$ where $\mathbb{P}_{\theta}^{\prime}:= \mathbb{P}_{\theta} \circ S^{-1}$.

In asymptotics one often  needs to show that a sequence of models converges to a limit model without being statistically equivalent to it at any point. This can be formulated by using LeCam's notion of deficiency and the associated distance on the space of statistical models. The deficiency of $\mathcal{P}$ with respect to $\mathcal{P}^{\prime}$ (expressed here in $L^{1}$ rather than total variation norm) is
$$
\delta(\mathcal{P},\mathcal{P}^{\prime}) := \inf_{R} \sup_{\theta\in \Theta} \|R(p_{\theta}) - p_{\theta}^{\prime}\|_{1} 
$$
where the infimum is taken over all randomisations  $R$. The LeCam distance between $\mathcal{P}$ and  $\mathcal{P}^{\prime}$ 
is defined as 
$$
\Delta(\mathcal{P},\mathcal{P}^{\prime}):= 
\mathrm{max} ( \delta(\mathcal{P},\mathcal{P}^{\prime}) ,\delta(\mathcal{P}^{\prime},\mathcal{P}) ) ,
$$ 
and is equal to zero if and only if the models are equivalent. A sequence of models $\mathcal{P}^{(n)}$ converges strongly to $\mathcal{P}$ if 
$$
\lim_{n\to\infty}\Delta (\mathcal{P}^{(n)} , \mathcal{P}) = 0. 
$$
This can be used to prove the convergence of optimal procedures and risks for statistical decision problems. We illustrate this with the example of local asymptotic normality (LAN) for i.i.d.\ parametric models, whose quantum extension provides an alternative route  to optimal estimation in quantum statistics. Suppose that $\mathcal{P}$ is a model over an open set $\Theta\subset \mathbb{R}^{k}$ and that $p_{\theta}$ depends sufficiently smoothly on $\theta$ (e.g., $p_{\theta}^{1/2}$ is differentiable in quadratic mean), and consider the local i.i.d.\ models around $\theta_{0}$ with local parameter $h\in \mathbb{R}^{k}$
$$
\mathcal{P}^{(n)}:= \{ \mathbb{P}_{\theta_{0} +h/\sqrt{n}}^{n} :  \| h \| \leq C \}. 
$$  
LAN means that $\mathcal{P}^{(n)}$ converges strongly to the Gaussian shift model consisting of a single sample from an 
$k$-variate normal distribution with mean $h$ and variance  equal to the inverse Fisher information matrix of the original model at $\theta_{0}$
$$
\mathcal{N}:= \left\{ N(h , I_{\theta_{0}}^{-1})  : \| h \| \leq C\right\}.
$$

\subsection{Convergence of quantum statistical models}\label{sec.conv.quantum.models}
As we have seen, an important problem in quantum statistics is to find the most informative measurement for a given quantum statistical model and a given decision problem. A partial solution to this problem is provided by the quantum 
Cram\'{e}r-Rao theory which aims to construct lower bounds to the quadratic risk of any estimator, expressed solely in terms of the properties of the quantum states.  Classical mathematical statistics suggests that rather than searching for optimal decisions, more insight could be gained by analysing the structure of the quantum statistical models themselves, beyond the notion of quantum Fisher information. Therefore we will start by addressing a more basic question of how to decide whether two quantum models over a parameter space $\Theta$ are statistically equivalent,  or close to each other in a statistical sense. To answer this question we will introduce the notion of quantum channel, which is a  transformation of quantum states that could --  in principle -- be physically implemented in a lab, and should be seen as the analog of a classical randomisation which defines a particular data processing procedure.  The simplest example of such transformation is a unitary channel which rotates a state ($d\times d$ density matrix $\rho$) by means of a $d\times d$ unitary matrix $U$, i.e.,  
$$
\mathcal{U} :\rho \mapsto U\rho U^{*}.
$$
Since $\mathcal{U}$ can be reversed by applying the inverse unitary $U^{-1}$, we anticipate that it will map any quantum model into an equivalent one. More generally, a quantum channel $C:M(\mathbb{C}^{d})\to M(\mathbb{C}^{k})$ must satisfy the minimal requirement of being positive and trace preserving linear map, i.e., it must transform quantum states into quantum states in an affine way, similarly to the action of a classical randomisation. However, unlike the classical case, it turns out that this condition needs to be strengthened to the requirement that  $C$ is {\it completely positive}, i.e., the amplified maps 
$$
C\otimes {\rm Id}_{n} : M(\mathbb{C}^{d}) \otimes M(\mathbb{C}^{n}) \to M(\mathbb{C}^{d}) \otimes M(\mathbb{C}^{n}) 
$$
must be positive for all $n\geq 0$, where ${\rm Id}_{n} $ is the identity transformation on  $M(\mathbb{C}^{n})$. An example of a positive but not completely positive, and hence unphysical transformation, is the transposition 
$tr:  M(\mathbb{C}^{d}) \to  M(\mathbb{C}^{d}) $ with respect to a given basis. Indeed, the reader can verify that 
applying $tr\otimes {\rm Id}_{d}$ to any {\it pure entangled} state ( i.e., not a product state 
$|\psi\rangle \langle \psi| \otimes |\phi\rangle \langle \phi |$) produces a matrix which is not positive, hence not a state.
   
\begin{definition}
A linear map $C:M(\mathbb{C}^{d})\to M(\mathbb{C}^{k})$ which is completely positive and trace preserving is called a quantum channel.
\end{definition}

The Stinespring-Kraus Theorem \cite{Nielsen&Chuang} says a linear map $C:M(\mathbb{C}^{d})\to M(\mathbb{C}^{k})$ is completely positive map if and only if it is of the form 
$$
C(\rho)  = \sum_{i=1}^{dk} K_{i} \rho K_{i}^{*},
$$  
with $K_{i}$ linear transformations from $\mathbb{C}^{d}$ to $\mathbb{C}^{k}$, some of which may be equal to zero. Moreover, $C$ is trace preserving if and only if $\sum_{i} K_{i}^{*} K_{i}= \mathbf{1}_{d}$. In particular, if  the sum consists of a single non-zero term $V\rho V^{*}$, the action of the channel $C$ is to embed the state $\rho$ isometrically into a the $d$-dimensional subspace ${\rm Ran}(V) \subset\mathbb{C}^{k}$. As in the unitary case, it is easy to see that this action is reversible (hence noiseless) and maps any statistical model into an equivalent one. We are now ready to define the notion of equivalence of statistical models, as an extension of the classical characterisation. 
\begin{definition}\label{def.equivalence}
Let $ \mathcal{Q}:=\{\rho(\theta) \in M(\mathbb{C}^{d}):\theta\in\Theta\}$ and 
$\mathcal{R}:=\{\varphi(\theta)\in M(\mathbb{C}^{k}) : \theta\in\Theta\}$ be two quantum statistical models over $\Theta$. Then  $\mathcal{Q}$ is statistically equivalent to  $\mathcal{R}$  if there exist quantum channels 
$T:M(\mathbb{C}^{d}) \to M(\mathbb{C}^{k})$ and  $S:M(\mathbb{C}^{k}) \to M(\mathbb{C}^{d})$ such that for all $\theta\in\Theta$
$$
T(\rho( \theta))  = \varphi(\theta) \qquad \textrm{and}\qquad S(\varphi(\theta)) =  \rho( \theta).
$$
\end{definition}
The interpretation of this definition is immediate. Suppose that we want to solve a statistical decision problem concerning the model $\mathcal{R}$, e.g., estimating $\theta$, and we perform a measurement $M$ on the state $\varphi_{\theta}$ whose outcome is the estimator $\hat{\theta}$ with distribution $\mathbb{P}^{M}_{\theta} = M(\rho(\theta))$ and risk $R^{M}_{\theta}:=\mathbb{E}_{\theta}( d(\hat{\theta} ,\theta)^{2})$. Consider now the same problem for the model 
$\mathcal{Q}$, and define the measurement $N= M \circ R$ realised by first mapping the quantum states $\rho(\theta)$ through the channel $T$ into $\varphi(\theta)$, and then performing the measurement $M$. Clearly, the distribution of the obtained outcome is again $\mathbb{P}^{M}_{\theta}$ and the risk is $R^{M}_{\theta}$, so we can say that $\mathcal{Q}$ is at least as informative as $\mathcal{P}$ from a statistical point of view. By repeating the argument in the opposite direction we conclude that any statistical decision problem is equally difficult for the two models, and hence they are equivalent in this sense. However, unlike the classical case the opposite implication is not true. For instance, models whose states are each other's transpose have the same set of risks for any decision problem but are usually not equivalent in the sense of being connected by quantum channels.  It turns out that a full statistical interpretation of Definition \ref{def.equivalence} is possible if one considers a larger set of {\it quantum decision problems}, which do not involve measurements, but quantum channels as statistical procedures.

Until this point we have tacitly assumed that any (finite dimensional) quantum model is built upon the algebra of square matrices of a certain dimension. However this setting is too restrictive as it excludes the possibility of considering hybrid classical-quantum models, as well as the development of a theory of quantum sufficiency. We motivate this extension through the  following example. We throw a coin whose outcome $X$ has probabilities $p_{\theta}(1)=\theta$  and $p_{\theta}(0)=1-\theta$, and subsequently we prepare a quantum system in the state $\rho_{\theta}(X) \in M(\mathbb{C}^{d}) $ which depends on $X$ and the parameter $\theta$. What is the corresponding statistical model ? Since the ``data'' is both classical and quantum, the ``state'' is a matrix valued density on $\{0,1\}$
$$
\varrho_{\theta}(i) = p_{\theta}(i) \rho_{\theta}(i) , \qquad i\in \{0,1\}
$$
or equivalently, a block-diagonal density matrix $\varrho_{\theta}(1)\oplus\varrho_{\theta}(2) \in M(\mathbb{C}^{d}) \oplus M(\mathbb{C}^{d})$ which is positive and normalised in the usual sense. While this can be seen as a state on the full matrix algebra $M(\mathbb{C}^{2d})$, it is clear that since the off-diagonal blocks have expectation zero for all $\theta$, we can restrict $\varrho_{\theta}$ to the block diagonal sub-algebra $M(\mathbb{C}^{d}) \oplus M(\mathbb{C}^{d})$ without loosing any statistical information. In other words, the latter is a sufficient algebra of our quantum statistical model. In general, for a model defined on some matrix algebra, one can ask what is the smallest sub-algebra to which we can restrict without loosing statistical information, i.e., such that the restricted model is equivalent to the original one in the sense of definition \ref{def.equivalence}. The theory of quantum sufficiency was developed in \cite{Petz&Jencova} where a number of classical results were extended to the quantum set-up, in particular the fact that the minimal sufficient algebra is generated by the likelihood ratio statistic.

We now make a step further and characterise the ``closeness'' rather than equivalence of quantum statistical models, by generalising LeCam's notion of deficiency between models.

\begin{definition}\label{def.lecam.distace}
Let $ \mathcal{Q}:=\{\rho(\theta) \in M(\mathbb{C}^{d}):\theta\in\Theta\}$ and 
$\mathcal{R}:=\{\varphi(\theta)\in M(\mathbb{C}^{k}) : \theta\in\Theta\}$ be two quantum statistical models over $\Theta$. 
The deficiency of $ \mathcal{R}$  with respect to $ \mathcal{Q}$ is defined as 
\begin{equation}\label{eq.deficiency}
\delta(\mathcal{R}, \mathcal{Q}) = \inf_{T} \sup_{\theta\in \Theta} \| \varphi(\theta) -  T (\rho(\theta))\|_{1}
\end{equation}
where the infimum is taken over all channels $ T:M(\mathbb{C}^{d})\to M(\mathbb{C}^{k})$. The LeCam distance between 
$\mathcal{Q}$ and $\mathcal{R}$ is 
$$
\Delta(\mathcal{Q}, \mathcal{R}) = \mathrm{max} \left( \delta(\mathcal{R}, \mathcal{Q}) , \delta(\mathcal{Q}, \mathcal{R}) \right).
$$
\end{definition}
This is an extension of the classical definition of deficiency for dominated statistical models. We will use the LeCam distance to formulate the concept of local asymptotic normality for quantum states and find asymptotically optimal measurement procedures.

\subsection{Continuous variables systems and quantum Gaussian states}

In this section we introduce the basic concepts associated to  continuous variables (cv) quantum systems, and  
then analyse the problem of optimal estimation for simple quantum Gaussian shifts models. 

Firstly we will restrict our attention to the elementary ``building block'' cv system which physically may be a particle moving on the real line, or a mono-chromatic light pulse. Then we will show how more complex cv systems can be reduced to a tensor product of such ``building blocks''  by a standard ``diagonalisation'' procedure.  

The Hilbert space of the system is $\mathcal{H}  =  L^{2}(\mathbb{R})$ and its quantum states  are given by density matrices, i.e., positive operators of trace one. Unlike the finite dimensional case, their linear span, called the space of trace-class operators $\mathcal{T}_{1}(\mathcal{H})$, is a proper subspace of all bounded operators on $\mathcal{H}$, which is a Banach space with respect to the trace-norm 
$$
 \|\tau \|_{1} := {\rm Tr}(| \tau |)  = \sum_{i=1}^{\infty} s_{i},
$$ 
where $s_{i}$ are the singular values of $\tau$. The key observables are two ``canonical coordinates'' ${\bf Q}$ and ${\bf P}$ representing the position and momentum of the particle, or the electric and magnetic field of the light pluse, and are defined as follows
\begin{equation}\label{def.qp}
({\bf Q} f)(x) = x f (x),\qquad
({\bf P}f )(x) = -i \frac{d f}{dx}(x). 
\end{equation} 
Although they do not commute with each other, they satisfy Heisenberg's commutation relation which essentially captures the entire algebraic properties of the system:
$$
{\bf Q} {\bf P}-{\bf P}{\bf Q} = i\mathbf{1}.
$$
The label ``continuous variables'' stems from the fact that the probability distributions of ${\bf Q}$ and ${\bf P}$ are always absolutely continuous with respect to the Lebesgue measure. Indeed since any state is a mixture of pure states, it suffices to prove this for a pure state $|\psi\rangle\langle \psi |$. If $Q$ and  $P$ denote the real valued random variables representing the outcomes of measuring ${\bf Q}$ and respectively ${\bf P}$ then using \eqref{def.qp} one can verify that
\begin{eqnarray*}
\mathbb{E}(e^{i uQ}) &=& \langle \psi , e^{iu{\bf Q}} \psi\rangle = \int  e^{iuq} |\psi(q)|^{2}dq,\\
\mathbb{E}(e^{i vP}) &=&  \langle \psi , e^{iv{\bf P}} \psi\rangle = \int  e^{ivp} |\widehat{\psi}(p)|^{2}dp.
\end{eqnarray*}
where $\widehat{\psi}$ is the Fourier transform of $\psi$. This means that $Q$ and $P$ have probability densities 
$|\psi(q)|^{2}$ and respectively $| \widehat{\psi}(p)|^{2}$, and suggests that the cv system should be seen as the non-commutative analogue of an $\mathbb{R}^{2}$ valued random variable. Following up on this idea we define the ``quantum characteristic function'' of a state $\rho$
$$
\widetilde{W}_{\rho}(u,v):= {\rm Tr}\left(\rho e^{-i (u{\bf Q} + v{\bf P})}\right)
$$
and the  Wigner or ``quasidistribution'' function
$$
W_{\rho}(q,p) =\frac{1}{(2\pi)^{2}}\int\!\int e^{i( uq+ vp)} \,  \widetilde{W}_{\rho}(u,v) du \, dv.
$$
These functions have a number of interesting and useful properties, which make them into important tools in visualising and analysing states of cv quantum systems. 
\begin{enumerate}
\item
there is a one-to-one correspondence between $\rho$ and $W_{\rho}$;
\item
the Wigner function may take negative values, but its marginal along any direction $\phi$ is a bona-fide probability 
density corresponding to the measurement of the quadrature observable ${\bf X}_{\phi}:= {\bf Q} \cos \phi +{\bf P} \sin\phi$;
\item
Both $W_{\rho}$ and $\widetilde{W}_{\rho}$ belong to $L^{2}(\mathbb{R}^{2})$ and the following isometry holds between the space of Hilbert-Schmidt operators $\mathcal{T}_{2}(L^{2}(\mathbb{R}))$ and  $L^{2}(\mathbb{R}^{2})$
$$
{\rm Tr}( \rho A) = \int\!\int W_{\rho} (q,p) W_{A} (q,p) \, dq \, dp.
$$
\end{enumerate}

We can now introduce the class of quantum Gaussian states by analogy to the classical definition.
\begin{definition}
Let $\rho$ be a state with mean $(q,p)= ({\rm Tr}(\rho{\bf Q}) ,{\rm Tr}(\rho{\bf Q}))$ and covariance matrix
$$
V:=
\left(
\begin{array}{cc}
{\rm Tr }\left(\rho (Q-q)^{2}\right)    &  {\rm Tr }\left(\rho  (Q-q)\circ (P-p)\right)\\
&\\
{\rm Tr }\left( \rho  (Q-q)\circ (P-p)\right) &  {\rm Tr }\left(\rho  (P-p)^{2}\right)
\end{array}
\right).
$$
Then $\rho$ is called Gaussian if its characteristic function is 
$$
{\rm Tr}\left(\rho e^{-i (u{\bf Q} + v{\bf P})}\right) = 
e^{-i t x^{t} } \cdot e^{-  t V t^{t}  /2  } , \qquad t= (u,v) , ~~x= (q,p),
$$
in particular the Wigner function $W_{\rho}$ is equal to the probability density of $N(x, V)$. 
\end{definition} 

While the definition looks deceptively similar to that of a classical normal distribution, there are a couple of important differences. The first one is that the covariance matrix $V$ cannot be arbitrary but must satisfy the {\it uncertainty principle}
\begin{equation}\label{eq.uncertainty}
{\rm Det} (V) \geq \frac{1}{4}.
\end{equation}
This restriction can be traced back to the commutation relations $[{\bf Q},{\bf P}]=i{\bf 1}$ which says that we cannon assign classical values to ${\bf Q}$  and ${\bf P}$ simultaneously. Which leads us to the second point, and the problem of optimal estimation: since ${\bf Q}$  and ${\bf P}$ cannot be measured simultaneously, their  covariance matrix $V$ is not ``achievable'' by any measurement aimed at estimating the means $(q,p)$ and the experimenter needs to make a trade-off between measuring ${\bf Q}$ with high accuracy  but ignoring ${\bf P}$, and vice-versa. In the last part of this section we look at this problem in more detail and explain the optimal measurement procedure. 

\begin{definition}
A quantum Gaussian shift model is family of Gaussian states 
$$
\mathcal{G} := \{\Phi(x,V): x\in \mathbb{R}^{2}\}
$$ 
with unknown mean $x$ and fixed and known covariance matrix $V$. 
If $G$ is a $2\times 2$ positive real weight matrix, the optimal estimation problem is to find the measurement 
$M$ with outcome $ \hat{x} = (\hat{q},\hat{p})$ which minimises the maximum quadratic risk
\begin{equation}\label{eq.quadratic.risk}
R(M) = \sup_{x} \mathbb{E}_{x} \left( ( \hat{x}-x) G (\hat{x}-x)^{t}  \right) .
\end{equation}
%
\end{definition}
This is a provisional definition only: a definitive version follows as Definition 6 below.
Finding the optimal measurement, relies on the equivariance (or covariance in physics terminology) of the problem with respect to the action of the translations (or displacements) group $\mathbb{R}^{2}$ on the states  
$$
\mathcal{D}(y) : \Phi(x,V) \mapsto \Phi(x+y, V) ,\qquad y\in \mathbb{R}^{2}.
$$
This action is implemented by a unitary channel 
$$
\Phi(x+y, V) = D(y) \Phi(x,V) D(y)^{*} ,\qquad  y=(u,v)
$$
where $D(y) = \exp(i v{\bf Q}- iu{\bf P} )$ are called the displacement or Weyl operators. Since $R(M)$ is invariant under the transformation $[x,\hat{x}] \mapsto [x+y , \hat{x}+y]$, a standard equivariance argument shows that the infimum risk is achieved on the special subset of {\it covariant} measurements, defined by the property
$$
\mathbb{P}_{\Phi(x+y,V)}^{(M)} (d\hat{x}+y) =  \mathbb{P}_{\Phi(x,V)}^{(M)} (d\hat{x}).
$$
Such measurements, and the more general class of covariant quantum channels, have a simple description in terms of linear transformation on the space of coordinates of the system together with an auxiliary system,  
 \cite{Werner&Nachtergaele}. More specifically, consider an independent quantum cv system with coordinates $({\bf Q}^{\prime}, {\bf P}^{\prime})$, prepared in a state $\tau$ with zero mean and covariace matrix $Y$.  By the commutation relations, the observables  ${\bf Q}+{\bf Q}^{\prime}$ and ${\bf P}-{\bf P}^{\prime}$ commute with each other and hence can be measured simultaneously. Since the joint state of the two independent systems is $\Phi(x,V)\otimes \tau$, the outcome $(\hat{q},\hat{p})$ of the measurement is an unbiased estimator of $(q,p)$ with covariance matrix $V+Y$, and the risk is 
 $$
 R(M) = {\rm Tr}(G (V+Y))=  {\rm Tr}(G V)+  {\rm Tr}(G Y)
 $$
 where the first term is the risk of the corresponding classical problem, and the second is the non-vanishing contribution due to the auxiliary ``noisy'' system.  To find the optimum, it remains to minimise the above expression over all possible covariance matrices of the auxiliary system which must satisfy the constraint ${\rm Det}(Y) \geq 1/4$. If $G$ has the form $G= O \,{\rm Diag}(g_{1},g_{2})\, O^{t}$ with $O$ orthogonal, then it can be easily verified that the optimal $Y$ is the matrix 
$$
Y_{0} = 
\frac{1}{2} O  \left(
\begin{array}{cc}
\sqrt{g_{2}/g_{1}}    & 0 \\
0 & \sqrt{g_{1}/ g_{2}}  \\
\end{array}
\right)O^{t}.
$$
Moreover, the unique state with such ``minimum uncertainty'' is the Gaussian state $\tau = \Phi(0,Y_{0})$. In conclusion, the minimax risk is
$$
R_{minmax}= \inf_{M}R(M) = {\rm Tr}(GV) + \sqrt{{\rm Det}(G)}.
$$ 

\subsection{General Gaussian shift models and optimal estimation}
\label{sec.general.gaussian}
We now extend the findings of the previous section from the ``building block'' system to a multidimensional setting. In essence, we show that the Holevo bound is \emph{achievable} for general Gaussian shift models, a result which has been known -- in various degrees of generality -- since the pioneering work of V.P. Belavkin and of A.S. Holevo in the 70's. 

Let us consider a system composed of $p\geq 1$ mutually commuting pairs of canonical coordinates $({\bf Q}_{i}, {\bf P}_{i})$, so that the commutation relations hold
 $$
 [{\bf Q}_{i}, {\bf P}_{j}] = i\delta_{i,j} \mathbf{1} , \qquad i,j=1,\dots, p.
$$
The joint system can be represented on the Hilbert space $L^{2}(\mathbb{R})^{\otimes p}$ such that the pair $({\bf Q}_{i}, {\bf P}_{i})$ acts on $i$-th copy of the tensor product as in \eqref{def.qp}, and as identity on the other spaces. Additionally, we allow for a number $l$ of ``classical variables'' 
${\bf C}_{k}$ which commute with each other and with all $({\bf Q}_{i}, {\bf P}_{i})$, and can be represented separately as position observables on $k$ additional copies of $L^{2}(\mathbb{R})$. For simplicity we will denote  all variables as 
$$
({\bf X}_{1},\dots, {\bf X}_{m})\equiv ({\bf Q}_{1}, {\bf P}_{1},\dots ,{\bf Q}_{p}, {\bf P}_{p}, {\bf C}_{1},\dots, {\bf C}_{l}),\quad m=2p+l,
$$
and write their commutation relations as
$$
[{\bf X}_{i}, {\bf X}_{j}] = iS_{i,j} \mathbf{1},
$$
where $S$ is the $m \times m$ block diagonal symplectic matrix of the  form 
$S= {\rm Diag}(\Omega, \dots ,\Omega,0,\dots 0 )$ with 
$$
\Omega= \left(
\begin{array}{cc}
  0&1\\
-1 &0
 \end{array}
 \right).
$$

Note that while this may seem to be rather special cv system, it actually captures the general situation since any symplectic (bilinear antisymmetric) can be transformed into the above one by a change of basis.

The states of this hybrid quantum-classical system are described by positive normalised densities in 
$\mathcal{T}_{1}(L^{2}(\mathbb{R}^{p} ))\otimes L^{1}(\mathbb{R}^{l})$, e.g., if the quantum and classical variables are independent the state is of the form $\rho\otimes p$ with $\rho$ a density matrix and $p$ a probability density. In general the classical and quantum parts may be correlated, and the state is a positive operator valued density 
$\varrho: \mathbb{R}^{l}\to \mathcal{T}_{1}(L^{2}(\mathbb{R}^{p} ))$, whose characteristic function can be computed as
$$
\mathbb{E}_{\varrho} \left(e^{i \sum_{i=1}^{m} u_{i}{\bf X}_{i} }\right) = 
\int\!\!\dots\!\int 
{\rm Tr }\left( \varrho(y) e^{ \sum_{i=1}^{2p}u_{i} {\bf X}_{i}} \right)  \, e^{i \sum_{j=1}^{l} u_{2p+j} y_{j} } \,dy_{1}\dots dy_{l}.
 $$
\begin{definition}
 A state $\Phi(x,V)$ with mean $x\in \mathbb{R}^{m}$ and  $m\times m$ covariance matrix $V$ is Gaussian if
$$
\mathbb{E}_{\Phi(x,V)} \left(e^{i \sum_{i=1}^{m} u_{i}{\bf X}_{i} }\right) = e^{i u x^{t} } e^{-   uVu^{t}/2 }. 
$$ 
A Gaussian shift model over the parameter space $\Theta:=\mathbb{R}^{k}$ is a family 
$$
\mathcal{G} := \{ \Phi (L h , V) :h\in \mathbb{R}^{k}\}
$$
where $L:\mathbb{R}^{k} \to\mathbb{R}^{m}$ is a linear map. 
\end{definition}
Note that the dimension of the parameter $h$ may be smaller than the dimension of mean value $x$. One may distinguish \emph{full} and \emph{partial} quantum Gaussian shift models: in the full model case, the dimensions are equal (and the matrix $L$ invertible). A non-classical feature of the general quantum Gaussian shift is that \emph{a linear submodel of a full Gaussian shift model is not, in general, equivalent to a full model with lower-dimensional mean vector}. 

The analogue of the uncertainty principle \eqref{eq.uncertainty}  for general cv systems is 
the (complex) matrix inequality 
\begin{equation}\label{eq.matrix.ineq.uncertainty}
V\geq  \frac{i}{2} S.
\end{equation}

The statistical decision problem is to find the  measurement which optimally estimates the parameter $h$ of the Gaussian state $\Phi(Lh, V)$, for a mean square error risk with a  given $k\times k$ weight matrix $G$, cf.\ \eqref{eq.quadratic.risk}. As before, we can restrict our attention to covariant measurements, i.e., to measuring {\it mutually commuting} variables of the form
$$
{\bf W}^{(i)} ={\bf  Y}^{(i)} + \tilde{\bf Y}^{(i)}
$$
where 
$$
{\bf Y}^{(i)}= \sum_{j=1}^{m} y^{(i)}_{j} {\bf X}_{j} , \qquad \mathbb{E}_{\Phi(Lh,V)} ( {\bf Y}^{(i)}) = h_{i}
$$
and 
$$
\tilde{\bf Y}^{(i)}= \sum_{j=1}^{\tilde{m}} \tilde{y}^{(i)}_{j} \tilde{{\bf X}}_{j}, \qquad \mathbb{E}_{\varrho} ( \tilde{\bf Y}^{(i)}) =0.
$$
Here $(\tilde{{\bf X}}_{1}, \dots ,\tilde{{\bf X}}_{\tilde{m}})$ are the coordinates of an independent, auxiliary system with symplectic matrix $\tilde{S}$, prepared in a state $\varrho$ with mean zero and covariance matrix $\tilde{V}$. Let 
$V^{({\bf Y})}$ and $V^{(\tilde{\bf Y})}$ denote the covariance matrices of the independent systems 
$( {\bf Y}^{(1)} ,\dots, {\bf Y}^{(k)}) $ and $(\tilde{\bf Y}^{(1)}, \dots , \tilde{\bf Y}^{(k)})$. Then the risk of the $({\bf W}^{(1)} ,\dots , {\bf W}^{(k)} )$ measurement is 
$$
R({\bf W})= {\rm Tr}(G V^{({\bf Y})}) + {\rm Tr}(G V^{(\tilde{\bf Y})}).
$$
On the other hand, since all ${\bf W}^{(i)}$ must commute with each other, we have
$$
[ \tilde{\bf Y}^{(i)} ,  \tilde{\bf Y}^{(j)}] = - [ {\bf Y}^{(i)} ,  {\bf Y}^{(j)}]  := -i S^{({\bf Y})}_{i,j} \mathbf{1}.
$$
The uncertainty principle \eqref{eq.matrix.ineq.uncertainty} applied to to the auxiliary variables 
$\tilde{\bf Y}^{(i)}$ gives the constraint
$$
V^{(\tilde{\bf Y})}\geq \pm\frac{i}{2}S^{({\bf Y})} .
$$
\begin{lemma}
Let $V$ and $S$ be real symmetric and respectively anti-symmetric $k\times k$ matrices, such that $V\geq iS/2$. Then
$
{\rm Tr}(V) \geq {\rm Tr}(|S|)/2,
$ 
with equality for $V= |S|/2$.
\end{lemma}
By optimising $V^{(\tilde{\bf Y})}$'s contribution to the risk and applying the above lemma with a fixed choice of 
${\bf Y}^{(i)}$ we obtain
$$
\inf_{\tilde{\bf Y}^{(i)}}  {\rm Tr}(G V^{(\tilde{\bf Y})})= \inf_{\tilde{\bf Y}^{(i)}}  {\rm Tr}(\sqrt{G} V^{(\tilde{\bf Y})} \sqrt{G})=
\frac{1}{2}{\rm Tr}(\sqrt{G} \left| S^{({\bf Y})} \right|\sqrt{G}).
$$
and the infimum is achieved for the covariance matrix $V^{(\tilde{\bf Y})}= |S^{(\tilde{\bf Y})}|/2$, which is only possible if the auxiliary system is prepared in the Gaussian state 
$\Phi(0, V^{(\tilde{\bf Y})})$, \cite{Leonhard}.

It remains now to optimise the risk over all unbiased $({\bf Y}^{(1)},\dots ,{\bf Y}^{(k)})$ i.e., 
which satisfy the condition \eqref{eq.unbiasedness} from the formulation of the Holevo bound:
\begin{equation}\label{eq.unbiasedness.gaussian}
\frac{\partial}{\partial h_{j}} \mathbb{E}_{\Phi_{h,V}}\left( {\bf Y}^{(i)}\right)= \delta_{i,j}.
\end{equation}
The minimax risk is then 
$$
R_{minmax}(\mathcal{G},G)= \inf_{ \{{\bf Y}^{(i)}\}} {\rm Tr }\left(\sqrt{G} V^{({\bf Y})} \sqrt{G}\right)+ 
\frac{1}{2}{\rm Tr}\left(\sqrt{G} \left| S^{({\bf Y})} \right|\sqrt{G} \right)
$$
which is equal to the Holevo bound \eqref{defCG2} if we consider that
$$
V^{{\bf Y}}_{i,j}=\Re \, \mathbb{E}_{\Phi(0,V)} ({\bf Y}^{(i)} {\bf Y}^{(j)}), \quad \text{and}\quad 
\frac{1}{2}S^{({\bf Y})} =\Im \, \mathbb{E}_{\Phi(0,V)} ({\bf Y}^{(i)}{\bf Y}^{(j)}).
$$

\subsection{Local asymptotic normality for i.i.d.\ states}

In this section we show how the general Gaussian shift models discussed above emerge from i.i.d.\ models through local asymptotic normality.

Suppose that we are given $N$ independent quantum systems prepared identically in an unknown state $\rho\in M(\mathbb{C}^{d})$. For large $N$ we can sacrifice a small part of the systems (e.g., $\tilde{N}= N^{1-\epsilon}$)  and use them to construct an estimator $\rho_{0}$ of the state, by means of a quantum tomography procedure. Using standard concentration inequalities it can be shown that $\rho$ belongs to a neighbourhood of size $N^{-1/2 +\epsilon}$ centred at $\rho_{0}$, with probability converging to one. Therefore,  the asymptotic behaviour of parameter estimation problems is determined by the structure of local quantum models around a fixed state $\rho_{0}$, and from now on we will restrict our attention to such models. By choosing the eigenvectors of 
$\rho_{0}$ as the standard basis, and assuming that the eigenvalues satisfy $\mu_{1} >\dots \mu_{d}>0$, we have $\rho_{0}= {\rm Diag}(\mu_{1},\dots, \mu_{d})$ and  
an arbitrary state in its neighbourhood is of the form
\begin{equation}\label{rho.theta.tilde}
\rho_{h} 
:=
\begin{bmatrix} 
\mu_1 + u_1 & \zeta_{1,2}^* & \dots & \zeta_{1,d}^*
\\
\zeta_{1,2} & \mu_2  + u_2 & \ddots& \vdots 
\\
\vdots & \ddots & \ddots & \zeta_{d-1,d}^* \\
\zeta_{1,d} & \dots & \zeta_{d-1,d} &\mu_d - \sum_{i=1}^{d-1} u_i
\end{bmatrix},
\qquad   u_i\in \mathbb{R}, ~ \zeta_{j,k}\in \mathbb{C}.
\end{equation}
with local parameter $h= (\vec{u}, \vec{\zeta})\in \mathbb{R}^{d-1}\times \mathbb{C}^{d(d-1)/2}\cong \mathbb{R}^{d^{2}-1}$. The local i.i.d.\ quantum model around $\rho_{0}$ is then defined as
\begin{equation}\label{eq.qn}
\mathcal{Q}_{N} := \left\{\rho^{N}_{h}:= \rho_{h/\sqrt{N}}^{\otimes N} : \|h\|\leq N^{\epsilon}\right\}.
\end{equation}

If some eigenvalues $\mu_{i}$ are equal to one another or to zero, degeneracies occur which are tricky to deal with. Completing the theory for such situations is a topic of ongoing research. In the rest of this section we give an intuitive argument for the emergence of the limit Gaussian model 
and finish with the precise formulation of LAN, restricting attention to the nondegenerate situation. 

We define $m=d^{2}-1$ operators whose expectation with respect to the state $\rho_{0}$ is zero, and together with the identity form a basis of  of the space of selfadjoint $d\times d$ matrices
$$
\{X_{1},\dots, X_{m}\} = \{ Q_{1,2} , P_{1,2},\dots,  Q_{d-1,d} , P_{d-1,d} ,C_{1},\dots, C_{d-1}\},
$$
where
$$
 Q_{j,k} := \frac{|j\rangle \langle k| + |j\rangle \langle k|}{\sqrt{2(\mu_{j} -\mu_{k})}},
\quad 
P_{j,k}:=  \frac{ i(  |k\rangle \langle j| - |j\rangle \langle k|)}{\sqrt{2(\mu_{j} -\mu_{k})}},
\quad
C_{i}:= |i\rangle\langle i| -\mu_{i} \mathbf{1}.
$$ 
Let $Q_{j,k}(N)\in M(\mathbb{C}^{d})^{\otimes N}$ denote the corresponding collective observables 
$$
Q_{j,k}(N) := \sum_{s=1}^{N} Q_{j,k}^{(s)} ,\quad  ~~Q_{jk}^{(s)}:=
\mathbf{1}\otimes \dots \otimes Q_{j,k}\otimes \dots \otimes \mathbf{1},
$$
with $Q^{(s)}_{j,k}$ acting on the position $s$ of the tensor product; similar definitions hold for $P_{j,k}(N), C_{i}(N)$. 
The collective observables play the role of sufficient statistic for our i.i.d.\ model, and we would like to understand their asymptotic behaviour. Since all systems are independent and identically prepared, and the terms in each collective observable commute, we can apply classical Central Limit techniques to show that, under the state $\rho^{n}_{h}$, we have
\begin{eqnarray}
\frac{C_{i} (N) }{\sqrt{N}}   &\overset{\mathcal{L}}{\longrightarrow} & 
N\left(u_{i}, \mu_{i}(1-\mu_{i}) \right), \quad 
1\leq i \leq d-1; \nonumber \\
\frac{Q_{j,k} (N) }{\sqrt{N}}                    &\overset{\mathcal{L}}{\longrightarrow} & 
N\left(\Re\tilde{\zeta}_{j,k}, v_{j,k}\right)
\quad\qquad 1\leq j<k\leq d;
\nonumber\\
\frac{P_{j,k} (N) }{\sqrt{N}}                    &\overset{\mathcal{L}}{\longrightarrow} & 
N\left(\Im \tilde{\zeta}_{j,k}, v_{j,k}\right),
\quad\qquad 1\leq j<k\leq d, \nonumber
\end{eqnarray}
where $\tilde{\zeta}_{j,k}= \zeta_{j,k}/\sqrt{(\mu_{j} -\mu_{k})/2}$ and $v_{j,k}= 1/(2(\mu_{j} -\mu_{k}))$.
This indicates that the model converges to a Gaussian shift model, but does not tell us what the {\it covariance} and 
{\it commutation relations} of the different limit variables are. For this, we need a quantum CLT, that is a multivariate CLT which takes into account the fact that the collective variables do not commute with each other. Its precise formulation can be found in \cite{Ohya&Petz}, but for our purposes it is enough to give the following recipe. The limit is a general cv 
system as described in section \ref{sec.general.gaussian}, with $m=d^{2}-1$ coordinates 
$({\bf X}_{1},\dots, {\bf X_{m}}) = ({\bf Q}_{j,k}, {\bf P}_{j,k}, {\bf C}_{i})$  having the commutation relations
$$
[{\bf X}_{a}, {\bf X}_{b} ] = {\rm Tr}(\rho_{0} [ X_{a},X_{b}])\mathbf{1} = 2 i \Im {\rm Tr}(\rho_{0}  X_{a}X_{b})\mathbf{1},
$$
whose state is Gaussian with covariance matrix 
$$
V_{a,b} =  {\rm Tr}(\rho_{0}  (X_{a}X_{b}+X_{b}X_{a} )/2 ) = \Re{\rm Tr}(\rho_{0}  X_{a}X_{b})\mathbf{1}.
$$
It can be easily verified that thanks to our special choice of basis, $({\bf Q}_{j,k}, {\bf P}_{j,k})$ are pairs of position and momentum operators, which commute with all other coordinates and ${\bf C}_{i}$ are ``classical'' variables, cf.\ section \ref{sec.general.gaussian}. Moreover the covariance matrix is block diagonal, with each pair $({\bf Q}_{j,k}, {\bf P}_{j,k})$ having a $2 \times 2$ the covariance matrix $V^{\rm q}_{j,k}= v_{j,k}\mathbf{1}$, and no correlation with the other coordinates, and the classical variables have covariance matrix 
$$
V^{\rm cl}_{ij} := \delta_{ij} \mu_{i}  -\mu_{i}\mu_{j}, \qquad  i,j =1,\dots d-1.
$$
In summary, the limit Gaussian model consists of a tensor product between a Gaussian probability density and a density matrix of $d(d-1)/2$ independent quantum Gaussian states 
\begin{equation}\label{eq.gaussian.model}
G(h,\mu) := 
\mathcal{N}(u,V^{{\rm cl}}) \otimes \bigotimes_{j<k} \Phi\left( (\Re\tilde{\zeta}_{j,k}, \Im\tilde{\zeta}_{j,k}), \, V^{\rm q}_{j,k} \right).
\end{equation}

We can now formulate the LAN Theorem which shows that the i.i.d.\ model can be asymptotically approximated by the Gaussian one, by means of quantum-classical randomisations, as discussed in section \ref{sec.conv.quantum.models}. An alternative approach based on a generalisation of the notion of weak convergence of models, can be found in \cite{GJ}.
\begin{theorem}
\label{main}
Let $\mathcal{Q}_{N}$ be the i.i.d.\ quantum model \eqref{eq.qn} and let 
$$
\mathcal{G}_{N}:= \{ G(h,\mu) : \|h\|\leq N^{\epsilon}\}.
$$ 
be the Gaussian model with $G(h,\mu)$ defined in \eqref{eq.gaussian.model}. Then there exist channels (completely positive, normalised maps) 
\begin{eqnarray}
T_N &:& 
M(\mathbb{C} ^{d})^{\otimes N} \to L^{1}(\mathbb{R}^{d-1}) \otimes \mathcal{T}_{1}
\left(L^{2}(\mathbb{R})^{\otimes d(d-1)/2}\right),
\nonumber\\
S_N &:& L^{1}(\mathbb{R}^{d-1}) \otimes  \mathcal{T}_{1}
\left(L^{2}(\mathbb{R})^{\otimes d(d-1)/2}\right)\to M(\mathbb{C} ^{d})^{\otimes N},
\nonumber
\end{eqnarray}
such that 
$$
\lim_{N\to \infty}\Delta(\mathcal{Q} _N, \mathcal{G} _N) =0,
$$ 
where $\Delta(\cdot, \cdot)$ is the LeCam distance, cf.\ Definition \ref{def.lecam.distace}. 
\end{theorem}

Clearly, in the same i.i.d.\ setting, smooth lower-dimensional submodels of the model of a completely unknown state converge to a partial Gaussian shift model.

\subsection{Asymptotic attainability of the Holevo bound}

Besides its theoretical importance, local asymptotic normality has been used as a tool for solving various asymptotic  problems such as optimal quantum learning \cite{Guta&Kotlowski}, teleportation benchmarks \cite{Guta&Bowles}, quantum state purification \cite{Guta&Bowles2}. Here we give a short non-technical argument for the asymptotic attainability of the Holevo bound for i.i.d.\ models, using local asymptotic normality. 

In section \ref{sec.general.gaussian} we showed that the Holevo bound is attained for arbitrary classical-quantum Gaussian shift models. We then saw that the model of $N$ i.i.d.\ systems prepared in a completely unknown state converges locally to a  Gaussian shift model with $(d^{2}-1)$ parameters. If some prior information about the state of the systems is available, we consider a lower dimensional model $\rho_{\theta}\in M(\mathbb{C}^{d})$ with 
$\theta\in \Theta\subset \mathbb{R}^{k}$. By applying LAN to this sub-model of the ``full'' one, we find that it is approximated in the LeCam sense by a Gaussian shift of the form
$$
\mathcal{G}^{\prime}= \{G(Lh^{\prime}, \mu): h^{\prime}\in \mathbb{R}^{k}\} 
$$
where $L:\mathbb{R}^{k}\to \mathbb{R}^{d^{2}-1}$ is a linear map which depends only on the local behaviour of the restricted model around $\theta_{0}$. To identify the linear transformation $L$ we recall the correspondence between the collective variables and the limit continuous variables
$$
(Lh^{\prime})_{a}:= \mathbb{E}_{G(h^{\prime},\mu)}( {\bf X}_{a})= 
\lim_{N\to\infty}{\rm Tr}( \rho_{h^{\prime}}^{N} X_{a}(N)) =  
\sum_{i=1}^{k} h^{\prime}_{i}{\rm Tr}\left( \left .\frac{\partial \rho_{h^{\prime}}}{\partial h^{\prime}_{i}}  \right|_{h=0} X_{a}\right) 
$$
from which we deduce
$$
L_{i,a}={\rm Tr}\left( \left .\frac{\partial \rho_{h^{\prime}}}{\partial h^{\prime}_{i}}  \right|_{h=0} X_{a}\right). 
$$

By a technical but otherwise rather standard argument, one can show that the asymptotic minimax risk for the problem of estimating the local parameter $h^{\prime}$ converges  to the minimax risk 
for the same problem and the Gaussian model $\mathcal{G}^{\prime}$, where in both cases the loss function is quadratic with weight matrix $G$
$$
\lim_{N\to\infty} \inf_{M_{N}}\sup_{\| h^{\prime}\|\leq N^{\epsilon}} N R(M_{N},h^{\prime}) =R_{minmax}(\mathcal{G}^{\prime},G).
$$

The final step in proving the asymptotic attainability of the Holevo bound for finite dimensional systems it is to observe that its expression coincides with that of the minimax risk deduced in section \ref{sec.general.gaussian}, applied to the Gaussian shift model $\mathcal{G}^{\prime}$. The optimisation \eqref{defCG2} is performed over selfadjoint matrices satisfying the condition \eqref{eq.unbiasedness},  which becomes \eqref{eq.unbiasedness.gaussian} when translated into the cv language. Similarly, the real and imaginary parts of $Z(X)$ become the covariance and the symplectic matrices  
$V^{\bf Y}$ and respectively $S^{\bf Y}/2$.


{

\bibliographystyle{Chicago}

\raggedright



}

\newpage

\section*{Appendix: examples}


In the three examples discussed here, the loss function is
derived from a very popular (among the physicists) 
figure-of-merit in state estimation called
\emph{fidelity}.  Suppose we wish to estimate a state 
$\rho=\rho(\theta)$ by $\widehat\rho=\rho(\widehat\theta)$.
Fidelity measures the closeness of the
two states, being maximally equal to $1$ when the
estimate and truth coincide.
It is defined as $\mathrm{Fid}(\widehat\rho,\rho)=
\bigl(\mathrm{trace}(\sqrt {\rho^{\frac 12}\widehat \rho\rho^{\frac12}})\bigl)^2$
(some authors would call this \emph{squared} fidelity).
When both states are pure, thus $\rho=|\phi\rangle\langle\phi|$ and
$\widehat\rho=|\widehat\phi\rangle\langle\widehat\phi|$ where $\phi$ and $\widehat\phi$ are
unit vectors in $\mathbb C^d$, then $\mathrm{Fid}(\widehat\phi,\phi)
=|\langle \widehat\phi|\phi\rangle|^2$. There is an important characterization
of fidelity due to \citet{fuchs} which both explains its meaning and leads to many important
properties. Suppose $M$ is a measurement on the quantum system.
Denote by $M(\rho)$ the probability distribution of the outcome of
the measurement $M$ when applied to a state $\rho$. For two probability
distributions $P$, $\widehat P$ on the same sample space, let $p$ and
$\widehat p$ be their densities with respect to a dominating measure
$\mu$ and define the fidelity between these probability measures as
$\mathrm{Fid}(\widehat P,P)=\bigl(\int \widehat p^{\frac12} p^{\frac12}\mathrm d \mu\bigr)^2$.
In usual statistical language, this is the \emph{squared Hellinger affinity} between
the two probability measures. It turns out that
$\mathrm{Fid}(\widehat\rho,\rho)=\inf_M \mathrm{Fid}(M(\widehat \rho),M(\rho))$,
thus two states have small fidelity when there is a measurement which
distinguishes them well, in the sense that the Hellinger affinity between
the outcome distributions is small, or in other words, the $L_2$
distance between the root densities of the data under the two models
is large.  

Now suppose states are smoothly parametrized by a vector parameter $\theta$.
Consider the fidelity between two states with close-by parameter
values $\theta$ and $\widehat\theta$, 
and suppose they are measured with the same measurement $M$.
From the relation 
$ \int p^{\frac12}\widehat p^{\frac12}\mathrm d \mu = 1 -\frac 12 \| \widehat p^{\frac12} - p^{\frac12}\|^2$
and by a Taylor expansion to second order one finds
$1-\mathrm{Fid}(\widehat P,P)\approx\frac 14 (\widehat \theta -\theta)^\top
 I_M(\theta) (\widehat \theta -\theta)$ where $I_M(\theta)$ is the Fisher information in
 the outcome of the measurement $M$ on the state $\rho(\theta)$.
 We will define the \emph{Helstrom} quantum information matrix $H(\theta)$
 by the analogous relation 
\begin{equation}\label{helstrom}
1-\mathrm{Fid}(\widehat \rho,\rho)\approx\frac 14 (\widehat \theta -\theta)^\top
 H(\theta) (\widehat \theta -\theta).
 \end{equation}
It turns out that $H(\theta)$ is the smallest
 ``information matrix'' such that $I_M(\theta)\le H(\theta)$ for all measurements
 $M$.
 
Taking as loss function 
$l(\widehat\theta,\theta)=1-\mathrm{Fid}(\rho(\widehat \theta),\rho(\theta))$ we 
would expect (by a quadratic approximation to the loss) 
that $\mathbb E_\pi \EuScript C_{\frac14 H}$ is a
sharp asymptotic lower bound on $N$ times the Bayes risk. We will prove
this result for a number of special cases, in which by a fortuitous 
circumstance, the fidelity-loss function is \emph{exactly} quadratic
in a (sometimes rather strange) function of the parameter. 
The first two examples concern
a two-dimensional quantum system and are treated in depth in
\citet{baganetal05}; below we just outline some important
features of the application. In the second of those two examples
our asymptotic lower bound is an essential part of a proof of
asymptotic optimality of a certain measurement-and-estimation
scheme.

The third example concerns an unknown pure state of arbitrary
dimension. Here we are present a short and geometric 
proof of a surprising but little known  result of \citet{hayashi98} 
which shows that an extraordinarily simple measurement scheme
leads to  an asymptotically optimal estimator (providing the data
is processed efficiently). The analysis also
links the previously unconnected Holevo and Gill-Massar
bounds (\citealp{holevo82, gillmassar00}).

\subsection*{Example 1: Completely unknown spin half ($d$=2, $p$=3)}

Recall that a completely unknown $2$-dimensional quantum
state can be written 
$\rho(\theta)=\frac12(\mathbf 1 + \theta_1 \sigma_1+\theta_2\sigma_2
+\theta_3 \sigma_3)$, where $\theta$ lies in the unit ball in
$\mathbb R^3$. It turns out that  $\mathrm{Fid}(\widehat \rho,\rho)=
\frac 12 (1 + \widehat\theta \cdot \theta + (1-\|\widehat\theta\|^2)^{\frac12}(1-\|\theta\|^2)^{\frac12})$.
Define $\psi(\theta)$ to be the four-dimensional vector
obtained by adjoining $(1-\|\theta\|^2)^{\frac12}$ to $\theta_1$, $\theta_2$, $\theta_3$.
Note that this vector has constant length $1$.
It follows that  
$1-\mathrm{Fid}(\widehat \rho,\rho)=\frac14\|\widehat\psi-\psi\|^2$.
This is a quadratic loss-function for estimation of $\psi(\theta)$
with $\widetilde G = \mathbf 1$, the $4\times 4$ identity
matrix.
By Taylor expansion of both sides, we find that 
$\frac 14 H = \psi'^\top \widetilde G \psi' = G$
and conclude from Theorem 1 that $N$ times
$1-$ mean fidelity is indeed asymptotically lower bounded
by $\mathbb E_\pi \EuScript C_{\frac14 H}$.

In  \citet*{baganetal05} the exactly optimal measurement-and-estimation
scheme is derived and analysed in the case of a rotationally invariant
prior distribution over the unit ball. 
The optimal \emph{measurement} turns out not to
depend on the (arbitrary) radial part of the prior distribution,
and separates into two parts, one used for estimating 
the direction $\theta/\|\theta\|$, the other part for estimating
the length $\|\theta\|$. The Bayes optimal estimator of the length
of $\theta$ naturally depends on the prior. Because of these
simplifications it is feasible to compute the asymptotic value
of $N$ times the (optimal) Bayes mean fidelity, and this
value is $(3+2 \mathbb E_\pi\|\theta\|$)/4.

The Helstrom quantum information
matrix $H$ and the Holevo lower bound $\EuScript C_{\frac14 H}$ are also 
computed. 
It turns out that $\EuScript C_{\frac14 H}(\theta)=(3+2\|\theta\|)/4$.
Our asymptotic lower bound is not only correct but also, as expected,
sharp.

The van Trees approach does put some non-trivial conditions
on the prior density $\pi$. The most restrictive conditions are that
the density is zero at the boundary of its support and that the
quantity (\ref{defJpi}) be finite.
Within the unit ball everything is smooth, but there are
some singularities at the boundary of the ball. So our
main theorem does not apply directly to many priors
of interest. However there is an easy approximation
argument to extend its scope, as follows.

Suppose we start with a prior $\pi$ supported by the whole
unit ball which does not satisfy the conditions.  For any $\epsilon>0$
construct $\widetilde \pi=\widetilde \pi_\epsilon$ which is smaller than
$(1+\epsilon)\pi$ everywhere,  and $0$ for $\|\theta\|\ge 1-\delta$
for some $\delta>0$.
If the original prior $\pi$ is smooth enough we can arrange
that $\widetilde\pi$ satisfies the conditions of the van Trees
inequality, and makes (\ref{defJpi}) finite. $N$ times the Bayes risk
for $\widetilde\pi$ cannot exceed $1+\epsilon$ times 
that for $\pi$, and the same must also be true
for their limits. Finally, $\mathbb E_{\widetilde \pi_\epsilon}
\EuScript C_{\frac14 H}\to  \mathrm E_ \pi
\EuScript C_{\frac14 H}$ as $\epsilon\to 0$.

Some last remarks on this example: first of all, it is
known that \emph{only} collective measurements
can asymptotically achieve this bound. Separate measurements
on separate systems lead to strictly worse estimators.
In fact, by the same methods one can obtain the
sharp asymptotic lower bound $9/4$ (independent of the
prior), see Bagan, Ballester, Gill,
  Mu{\~n}oz-Tapia and Romero-Isart (2006b), when one allows
the measurement on the $n$th system to depend on
the data obtained from the earlier ones. Instead of
the Holevo bound itself, we use here a bound
of  \citet{gillmassar00}, which is actually has the form of
a dual Holevo bound.
(We give some more remarks on this at the end of the
discussion of the third example).
Secondly, our result gives strong heuristic support 
to the claim that the measurement-and-estimation
scheme developed in \citet*{baganetal05} 
for a specific prior and specific
loss function is also pointwise optimal in a minimax sense,
or among regular estimators, for loss functions which
are locally equivalent to fidelity-loss; and also asymptotically
optimal in the Bayes sense for other priors and locally
equivalent loss functions. In general, if the physicists' approach is
successful in the sense of generating a measurement-and-estimation
scheme which can be analytically studied and experimentally
implemented, then this scheme will have (for large $N$) good
properties independent of the prior and only dependent
on local properties of the loss.

\subsection*{Example 2: Spin half: equatorial plane ($d$=2, $p$=2)}

 \citet*{baganetal05}  also considered the case where
it is known that $\theta_3=0$, thus we now have a two-dimensional
parameter. The prior is again taken to be rotationally symmetric.
The exactly Bayes optimal measurement turns out (at least,
for some $N$ and for some priors)
to depend on the radial part of the prior. Analysis of
the exactly optimal measurement-and-estimation procedure
is not feasible since we do not know if this phenomenon persists
for all $N$. However there is a natural measurement,
which is exactly optimal for some $N$ and some priors, which
one might conjecture to be asymptotically optimal for all
priors. This sub-optimal measurement, combined with
the Bayes optimal estimator given the measurement,
can be analysed and it turns out that $N$ times
$1-$ mean fidelity converges to $1/2$ as $N\to \infty$,
independently of the prior. Again, 
the Helstrom quantum information
matrix $H$ and the Holevo lower bound $\EuScript C_{\frac14 H}$ are 
computed. 
It turns out that $\EuScript C_{\frac14 H}(\theta)=1/2$.
This time we can use our asymptotic lower bound to prove
that the natural sub-optimal measurement-and-estimator
is in fact asymptotically optimal for this problem.

For a $p$-parameter model the best one could every hope
for is that for large $N$ there are measurements with 
$\overline I_M$ approaching the Helstrom upper bound $H$.
Using this bound in the van Trees inequality gives the asymptotic
lower bound on $N$ times $1-$ mean fidelity of $p/4$.
The example here is a special case where this is attainable.
Such a model is called \emph{quasi-classical}.

If one restricts attention to separate measurements on
separate systems the sharp asymptotic lower bound is $1$,
twice as large, see Bagan, Ballester, Gill,
  Mu{\~n}oz-Tapia and Romero-Isart (2006b).

\subsection*{Example 3: Completely unknown $d$ dimensional pure state}

In this example we make use of the dual Holevo bound 
and symmetry arguments to show that in this example, the
original Holevo bound for a natural choice of $G$
(corresponding to fidelity-loss)
is attained by an extremely large class of measurements,
including one of the most basic measurements around,
known as ``standard tomography''.

For a pure state $\rho=|\phi\rangle\langle\phi|$, 
fidelity can be written $|\langle \widehat\phi|\phi\rangle|^2$
where $|\phi\rangle\in\mathbb C^d$ is a vector of unit length. 
The
state-vector can be multiplied by $e^{i a}$ for an arbitrary real phase
$a$ without changing the density matrix. The constraint of unit length
and the arbitrariness of the phase means that one can parametrize
the density matrix $\rho$ corresponding to $|\phi\rangle$ 
by $2(d-1)$ real parameters which we take to be our underlying
vector parameter $\theta$ (we have $d$ real parts and $d$ imaginary
parts of the elements of $|\phi\rangle$, but one constraint and one 
parameter which can be fixed arbitrarily).

For a pure state, $\rho^2=\rho$ so
$\mathrm{trace}(\rho^2)=1$.
Another way to write the fidelity in this case is as 
$\mathrm{trace}( \widehat\rho\rho)=\sum_{ij}(\Re(\widehat \rho_{ij})\Re(\rho_{ij})+
\Im(\widehat \rho_{ij})\Im(\rho_{ij}))$.  
So if we take $\psi(\theta)$ to be the vector
of length $2 d^2$ and of length $1$ containing the real and the imaginary parts of 
elements of $\rho$ we see that $1-\mathrm{Fid}(\widehat\rho,\rho)=\frac 12 \|\widehat\psi-\psi\|^2$.
It follows that $1-$ fidelity is a quadratic loss function in $\psi(\theta)$
with again $\widetilde G=\mathbf 1$. 

Define again the Helstrom quantum information matrix $H(\theta)$ for $\theta$
 by  $1-\mathrm{Fid}(\widehat \rho,\rho)\approx\frac 14 (\widehat \theta -\theta)^\top
 I_M(\theta) (\widehat \theta -\theta)$. Just as in the previous two examples
 we expect the asymptotic lower bound $\mathbb E_\pi \EuScript C_{\frac14 H}$
 to hold for $N$ times Bayes mean fidelity-loss, where $G=\frac14 H=
 \psi'^\top \widetilde G \psi' $.
 
 Some striking facts are known about estimation of a pure state. 
 First of all, from \citet{matsumoto02}, we know that the Holevo bound
 is attainable, for all $G$,  already at $N=1$. Secondly, from \citet{gillmassar00}
 we have the following inequality
 \begin{equation}\label{gillmassar}
 \mathrm{trace} H^{-1} \overline I_M\le d-1
 \end{equation}
 with \emph{equality} (in the case that the state is completely unknown) 
 for all \emph{exhaustive}
 measurements $M^{(N)}$ on $N$ copies of the state. Exhaustivity means,
 for a measurement with discrete outcome space, that $M^{(N)}(\{x\})$ 
 is a rank one matrix for
 each outcome $x$. The meaning of exhaustivity in general is by the same property for
 the density $m(x)$ of the matrix-valued measure $M^{(N)}$ with respect to a real
 dominating measure, e.g., $\mathrm{trace}(M^{(N)}(\cdot))$.
 This tells us that (\ref{gillmassar}) is one of the ``dual Holevo inequalities''.
 We can associate it with an original Holevo inequality once we know an 
 information matrix of a measurement attaining the bound. We will show that there
 is an information matrix of the form $\overline I_M=cH$ attaining the bound.
 Since the number of parameters (and dimension of $H$) is $2(d-1)$
 it follows by imposing equality in (\ref{gillmassar}) 
 that $c=\frac12$. The corresponding Holevo inequality 
 must be $ \mathrm{trace} \frac 12 H H^{-1} \frac 12 H \overline I_M^{-1}\ge d-1$
 which tells us that $\EuScript C_{\frac 14 H}=d-1$. 
 
 The proof uses an invariance property of the model. For any unitary matrix $U$
 (i.e., $UU^*=U^*U=\mathbf 1$) we can convert the pure state 
 $\rho$ into a new pure state $U\rho U^*$. The unitary matrices form a group
 under multiplication.  Consequently the group can be 
 thought to act on the parameter $\theta$ used to describe the pure state. Clearly the
 fidelity between two states (or the fidelity between their two parameters)
 is invariant when the same unitary acts on both states. This group action
 possesses the ``homogenous two point property'': for any two pairs of 
 states such that the fidelities between the members of each pair are the same, 
 there is a unitary transforming the first pair into the second pair.
 
 We illustrate this in the case $d=2$ where 
(first example, section 2), the pure
states can be represented by the surface of the unit ball in $\mathbb R^3$.
 It turns out that the action of the unitaries on the density matrices
 translates into the action of the group of orthogonal rotations on
 the unit sphere. Two points at equal distance on the sphere can
 be transformed by some rotation into any other two points at the
 same distance from one another; a constant distance between points
 on the sphere corresponds to a constant fidelity between the 
underlying states. 

In general, the pure states of dimension $d$ can be identified with
the Riemannian manifold $CP^{d-1}$ whose natural Riemannian
metric corresponds locally to fidelity (locally, $1-$ fidelity
is squared Riemannian distance) and whose isometries correspond
to the unitaries. This space posseses the homogenous two
point property, as we argued above. It is easy to show that the
\emph{only} Riemannian metrics invariant under isometries
on such a space are proportional to one another. Hence the
quadratic forms generating those metrics with respect to
a particular parametrization must also be proportional
to one another.

Consider a measurement whose outcome is actually an
estimate of the state, and suppose that this measurement
is \emph{covariant} under the unitaries. This means that
transforming the state by a unitary, doing the measurement
on the transformed state, and transforming the estimate back 
by the inverse of the same unitary, is the same 
(has the same POVM) as the original measurement.
The information matrix for such a measurement is generated
from the squared Hellinger affinity between the distributions
of the measurement outcomes under two nearby states,
just as the Helstrom information matrix is generated from
the fidelity between the states. If the measurement is
covariant then the Riemannian metric defined by the
information matrix of the measurement outcome must
be invariant under unitary transformations of the states.
Hence: \emph{the information matrix of any covariant measurement
is proportional to the Helstrom information matrix.}

Exhaustive covariant measurements certainly do exist. 
A particularly simple one is that, for each of the $N$ copies
of the quantum system, we independently and uniformly
choose a basis of $\mathbb C^d$ and perform the simple
measurement (given in an example in Section 2) 
corresponding to that basis.

The first conclusion of all this is: any exhaustive covariant
measurement has information matrix $\overline I_M^{(N)}$
equal to one half the Helstrom information matrix.
All such measurements attain the Holevo bound
$\mathrm{trace} \frac14 H ( \overline I_M^{(N)} )^{-1}\ge d-1$.
In particular, this holds for the i.i.d.\ measurement based
on repeatedly choosing a uniformly distributed random
basis of $\mathbb C^d$.

The second conclusion is that an asymptotic lower
bound on $N$ times $1-$ mean  fidelity is $d-1$. 
Now the exactly Bayes optimal measurement-and-estimation
strategy is known to achieve this bound. The measurement
involved is a mathematically elegant
collective measurement on the $N$ copies
together, but hard to realise in the laboratory. 
Our results show that one can expect to asymptotically
attain the bound by decent information processing
(maximum likelihood? optimal Bayes with uniform prior
and fidelity loss?)
following an arbitrary \emph{exhaustive covariant
measurement}, of which the most simple to implement
is the standard tomography measurement consisting of
an independent random choice of measurement basis
for each separate system.

In \citet{gillmassar00} the same bound as (\ref{gillmassar})
was shown to hold for separable (and in particular, for
adaptive sequential) measurements also in the mixed
state case. Moreover in the case $d=2$, any information
matrix satisfying the bound is attainable already at $N=1$.
This is used in \citet{baganetal06} to obtain sharp
asymptotic bounds to mean fidelity for separable
measurements on mixed qubits.

\end{document}